\documentclass{aastex}
\usepackage{emulateapj5}

\begin{document}

\def\lf{l_{\rm F}}
\def\tu{\tau_{\rm u}}
\def\VS{V\'azquez-Semadeni}

\title{The nonlinear development of the thermal instability in the
atomic ISM and its interaction with random fluctuations}
\author{F.J. S\'{a}nchez-Salcedo }

\affil{Instituto de Astronom\'\i a, UNAM, Ciudad Universitaria, Aptdo. 70 264,
C.P. 04510, Mexico City, Mexico}

\author{E. V\'azquez-Semadeni}
\and
\author{A. Gazol}
\affil{Instituto de Astronom\'\i a, UNAM, Campus Morelia, Aptdo. 72-3
(Xangari) 58089 Morelia,  
Michoac\'{a}n, Mexico}

\begin{abstract}

We discuss the nonlinear development of the isobaric mode of thermal
instability (TI) in the context of the atomic interstellar medium (ISM),
both in isolation and in the presence of either density or velocity
fluctuations, in order to assess the ability of TI to establish a
well-segregated multi-phase structure in the turbulent ISM.
The key parameter is the ratio of the cooling
time to the dynamical crossing time $\eta$.
First, we discuss the degree to which the condensation process of
large-scale perturbations generates large velocities, and the times
required for them to subside. Using high-resolution simulations 
in 1D and fits to recently published cooling rates, we find that
density perturbations of sizes 
$\gtrsim 15$ pc in media with mean density $\sim 1$ cm$^{-3}$ develop
inflow motions with Mach numbers larger than $0.5$ and a shock that propagates
outwards from the condensation, bringing the surrounding
medium out of thermal equilibrium. The time for the dynamical transient state
to subside ranges from 4 to 30 Myr for initial density perturbations of $20\%$
and sizes 3 to 75 pc. By the time the
condensations have formed, a substantial fraction of the mass is still
traversing the unstable range. Smaller (0.3 -- 3 pc) perturbations may
condense less dynamically, and reach nearly static configurations in shorter
times (e.g., $\sim 3.5$ Myr for perturbations of $\sim 0.3$ pc), but
they may be stable if they have a turbulent origin (see below). 
We thus suggest that, even if TI were the sole 
cloud-forming agent in the ISM,
clouds formed by it should be bounded by accreting gas traversing the 
unstable range, rather than by sharp transitions to the stable warm phase.
Second, we discuss the competition 
between a spectrum of density perturbations of various sizes. We
empirically find that, in order for small-scale perturbations not to
alter significantly the global evolution, progressively larger values of
$\eta$ are necessary as the initial spectrum becomes shallower. Finally,
we discuss the 
development of the instability in the presence of random velocity forcing,
which we argue is the most realistic way to emulate density fluctuation
production in the actual ISM. Such fluctuations are quasi-adiabatic
rather than quasi-isobaric in the large-$\eta$ limit, and trigger the
wave mode of TI, rather than the condensation mode, being stable to
first order. Indeed,
we find that {\it the condensation process can be suppressed for
arbitrarily long times} if the
forcing causes a moderate rms Mach number ($\gtrsim 0.3$) and extends to small
enough scales or occurs in low enough density environments that the turbulent
crossing time becomes smaller than the cooling time at those scales. 
We suggest that this mechanism, and the long times required to evacuate
the unstable phase, may be at the origin of the relatively large amounts of gas
mass in the unstable regime found in both observations and simulations
of the ISM. The gas with unstable temperatures is expected to be out of
thermal equilibrium, suggesting that it can be observationally
distinguished by simultaneously measuring two of its thermodynamic
variables. We remark that in the (stable) warm diffuse medium $\eta$
is large enough that the response to velocity perturbations of scales up 
to several parsecs is close to adiabatic, implying that it is relatively
weakly compressible, and thus consistent with recent observations that suggest
a nearly Kolmogorov power spectrum in this medium.

\end{abstract}


\section{Introduction}

The fact that the neutral atomic interstellar medium (ISM) is most
likely thermally bistable (Zeldovich \& Pikel'ner 1969; Field,
Goldsmith \& Habing 1969; Wolfire et 
al.\ 1995) has had a great impact on our picture of the formation of
interstellar structure. In particular, the unstable cooling process
has been thought to be responsible for the collection of mass into HI
clouds of sizes $\sim 1$ pc,
densities $\sim 10$--100 cm$^{-3}$, and temperatures $\sim 100$ K
(e.g., Goldsmith 1970; Schwarz, McCray \& Stein 1972; Parravano 1987;
Lioure \& Chi\`eze 1990; Koyama \& Inutsuka 2002). In the classical 
model of McKee \& Ostriker (1977), the cold and warm
neutral media are two distinct phases, with
clouds forming in dense shells created by shock compressions, 
aided by the thermal instability (TI hereafter), and later confined by
the warm medium's thermal pressure. The triggering of TI in the warm
medium by strong compressions and the subsequent segregation in two
separate phases has been recently studied by Hennebelle \&
P\'erault (1999, 2000) and by Koyama \& Inutsuka (2000). 

However, over the years, a more complex picture of the ISM has
emerged, driven in part by the advent of realistic numerical simulations
of the ISM that have highlighted the role of inertia and advection
(i.e., transport) by the velocity field. In this context, the role of the
thermal instability in the overall dynamics of the ISM is still
uncertain. Simulations of the turbulent ISM driven by supernovae  or
stellar winds suggest that the complex structure of the ISM 
as well as HI clouds can be the result of random gas 
motions, at least in regions with densities
below $10$--$100$ cm$^{-3}$, with no need for
thermal instabilities (e.g. Bania \& Lyon 1980; Rosen \& Bregman 1995;
Passot, V\'azquez-Semadeni \& Pouquet 1995; Korpi et al.\footnote{The 
simulations of Korpi et al.\ actually use a cooling function that  
imply a thermally unstable range between the cold and warm gas if no
heating is present (see, e.g., Murray \& Lin 1989; Burkert \& Lin 2000).
Although no global heating was
used in those simulations, the presence of local stellar heating causes
the gas to be thermally stable while subject to heating, and unstable
while cooling.} 1999; see also V\'azquez-Semadeni \& Passot 1999 and
V\'azquez-Semadeni 2002 for reviews; but see Wada, Spaans 
\& Kim 2000; Koyama \& Inutsuka 2002; Kritsuk \& Norman 2002 for the
alternative view that TI powers ISM turbulence).

Within this scenario, V\'azquez-Semadeni, Gazol \& Scalo 
(2000, hereafter Paper I) studied the probability density distribution
(PDF) of mass density in low resolution ($128^2$) 2D simulations of the
thermally bistable ISM at large scales (1-kpc box), finding that the
PDF may not reflect the thermally unstable 
nature of the gas in the presence of magnetic fields, Galactic shear,
the Coriolis force, and turbulence driven by stellar-like sources
located at the density maxima (clouds). Moreover, simulations at
higher resolution ($512^2$) with both the same box size (1 kpc; Gazol et al.\
2001, hereafter Paper II), and smaller ones (150 and 10 pc; \VS\ et
al. 2002, hereafter Paper III) have 
temperature histograms that indicate the existence of significant
amounts ($\sim 50\%$) of gas with temperatures in the 
unstable range. The same result could be seen in the simulations by
Korpi et al.\ (1999), although they did not discuss it, and has later
also been found in simulations of the late nonlinear stages of the pure
instability at smaller scales (6-pc box, at $256^3$ resolution) by
Kritsuk \& Norman (2002). This is 
in agreement with observations by Dickey, Salpeter \& Terzian (1978),
Kalberla, Schwartz, \& Goss (1985), Spitzer \& Fitzpatrick (1995),
Fitzpatrick \& Spitzer (1997), and Heiles (2001), as well as previous
theoretical suggestions (e.g., Lioure \& 
Chi\`eze 1990; Norman \& Ferrara 1996). 
Thus, the study of TI in
the context of fluctuating media becomes important, in particular to
understand the origin of the large fractions of gas mass in the unstable 
regime.

At the large scales studied by Paper I, the condensation\footnote{For
consistency with the nomenclature in the literature, we maintain the 
name ``condensation'' even for highly dynamic cases in which
``accretion'' or ``collapse'' might be more appropriate.}
process leading to the formation of clouds and intercloud ``voids''
becomes highly dynamic (i.e., develops large velocities in the nonlinear 
stage). The parameter controlling the degree to 
which the condensation process is dynamic is the ratio ($\eta$) of 
the cooling to the representative (sound or bulk) dynamical
time. Although it is 
customary in the literature to discuss TI in terms of the perturbation
wave lengths (or wavenumbers; e.g. Field 1965), in
this paper we prefer a description in terms of $\eta$  because it allows
the consideration of perturbations of a given physical size (say, in
parsecs), but that may correspond to different values of $\eta$ because
of, say, they occur at different mean densities (i.e., equilibrium
cooling rates).

It is well known (e.g., Field 1965; Murray \& Lin 1989;
Burkert \& Lin 2000) that perturbations with 
$\eta >1$ (corresponding to small-scale perturbations at a fixed cooling 
rate or low cooling rate at a fixed length scale)
evolve almost isobarically, and thus quasi-statically.
Instead, perturbations with $\eta <1$ develop large
pressure gradients, and thus evolve highly dynamically even under the
isobaric mode of the instability, although they take longer times to 
grow (e.g., Field 1965; Balbus 1995; Meerson 1996). 
Thus, Paper I suggested that large 
clouds formed by TI should be bounded by accretion shocks (rather than
by contact discontinuities), in a similar manner to that discussed by
David, Bregman \& Seab (1988) for TI in $10^6$--$10^8$-K gas, or by Rees
(1977) in the case of gravitational contraction. It is then
important to determine whether large, static, thermal
pressure-confined clouds ever
develop and, if yes, how long does it take for the dynamic situation
to fade away. If this time is long compared to estimated cloud
lifetimes [see Blitz \& Shu (1980) and Ballesteros-Paredes, Hartmann \&
\VS \, (1999) for two different possibilities in the case of molecular
clouds], or to the time scales for major disturbances to pass through 
the cloud (e.g., Kornreich \& Scalo 2000), then the static
configuration should not be expected to occur in general in the ISM.
Unfortunately, Paper I did not proceed to give a definitive solution 
to this question because of limitations of the numerical 
scheme used there. Most previous discussions of the time scales to reach  
a developed cloud-intercloud medium have been based on the assumption of
isobaric evolution (e.g., Parravano 1987), which,
however, is strictly valid only in the limit of $\eta \gg 1$. Goldsmith
(1970) did consider the limit $\eta \ll 1$, and mentioned that very long 
times are required for the quasi-static regime to be reached, such that his
simulations (at very low resolutions) never reached it (see also
Schwarz et al.~1972).

Small-scale perturbations, on the other hand, are good candidates for
rapid, yet not highly dynamic condensation, and constitute the 
paradigm of cloud formation via the isobaric mode of TI. 
This brings up the questions of a) whether, in the presence of a 
spectrum of density fluctuations, the small-scale perturbations can use up
most of the available mass before the large-scale ones grow, and thus
impede their development, and b) whether the paradigm is maintained when 
the perturbations are produced by turbulent velocity fluctuations. In
the regime of Paper I, in which the cooling times
are short, turbulence forced at large scales cannot prevent the
development of TI, and in fact actually enhances it, although
stellar-like heating at small spatial scales was shown to
erase the signature of TI in the density histogram. The latter result
does not strictly constitute an inhibition of TI, because the stars form 
{\it after} the condensations, which in turn formed by a combination of
the turbulent compressions and TI. Such condensations,
in the absence of the stellar heating,
would pile up in the thermally-stable density regime. However, true
avoidance of the condensation may occur for random velocity
fluctuations at the small scales because in this case the compressions
tend to heat the gas; i.e., the perturbations are quasi-adiabatic, and can
re-expand before they have time to cool.

The discussion above is related to the question of whether the
classical picture of cold, dense clouds in pressure equilibrium
with their warm, dilute surroundings is a realistic model of the ISM,
even if TI were the sole cloud-forming 
agent in the ISM (which it is not). 
Indeed, the equilibrium configuration is a possible solution of the 
hydrodynamic equations in the presence of TI-inducing cooling
functions; it is very often 
used as an initial condition of both analytical and numerical
studies, and even as a textbook case. 
However, the question is whether it is
also a self-consistent {\it final state} of previous evolution starting 
from small-amplitude perturbations of thermally unstable gas. Of
course, the pressure equilibrium configuration is valid as an initial 
condition for numerical simulations or as a case study for theoretical 
analyses, just as Gaussian random
fluctuations are in studies of turbulence or Cosmology, but they are
not to be interpreted as realistic representations of the ISM. For
example, shock-cloud collision simulations (e.g., Klein, McKee \&
Woods 1995) often start with a cloud 
in pressure equilibrium with its environment, but end up with fully
developed turbulence, suggesting that the latter is a more likely regime for
the ISM than their idealized initial conditions. 

In this paper we explore the nonlinear development of TI in the
isobaric mode, in the context of the atomic ISM, focusing on 
the issues mentioned above, using high resolution
numerical simulations in one dimension (1D) and a numerical scheme
better suited for this type of problems than that used in Paper I. 
We should emphasize that in this paper we choose to isolate the effects
of the instability, and concentrate on a
somewhat idealized case (one-dimensional geometry, simple
parameterized cooling functions neglecting non-equilibrium chemical
evolution, neglect of other physical agents such as the magnetic
field, differential disk rotation, self-gravity and realistic, stellar
energy injection), not necessarily representative of the full
situation in the ISM. We make no attempt to discuss the fate of the
clouds formed by TI in the real ISM, as this must be accomplished by
means of simulations of the fully turbulent ISM in the presence of the
agents mentioned above (see Papers I, II and III), and is
likely to involve dynamical instabilities whose study requires two- or
three-dimensional simulations (see, e.g., Murray et al.\ 1993; Vietri,
Ferrara \& Miniati 1997). Instead, our main goals here are simply to
illustrate, even in this simplified case, the following features of the
condensation process: a) the dynamical nature of the condensation
process, providing quantitative estimates of the perturbation sizes in
the atomic ISM required for such dynamical evolution to occur, and of
the times required for it to subside; b) the existence of accreting gas, 
which remains at densities in the unstable range,
even after the condensation process has reached a slowly evolving stage; 
c) the competition between growing modes of different sizes; d) the
evolution under the isobaric mode of TI of perturbations caused by
random velocity fluctuations, and e) their implications on standard
scenarios of 
a multiphase ISM in which most of the gas must reside in one of the
stable phases. 

The plan of the paper is as follows. We first present the
numerical method and discuss the resolution
requirements (\S \ref{sec:model}). Next, we present a qualitative
discussion of the timescales involved and the regimes that can appear
depending on the ratio $\eta$ (\S \ref{sec:quali}), to then consider
the evolution of Gaussian density perturbations, the size scales
at which shocks form at advanced stages of the
``condensation'' process, and the times required to reach quasi-static
states (\S \ref{sec:ev_density}). We then discuss the 
competition between the growth of small- and large-scale perturbations
by means of simulations with an initial spectrum of density 
fluctuations (\S \ref{sec:scale_compet}), and the evolution of turbulent 
fluctuations in the presence of TI by means of simulations including random
forcing (\S \ref{sec:vel_compet}). In \S \ref{sec:discussion} we
compare our results with previous work, and discuss the
limitations of the one-dimensional simulations we have used as well as
some implications of our results on the classical picture of 
thermal-pressure confinement of clouds, the existence of gas in the
unstable regime, and the compressibility of the warm gas in the ISM. Finally,
we summarize our results in \S \ref{sec:summary}.

\section{The model} \label{sec:model}

\subsection{Numerical method and relevant parameters} \label{sec:num_meth}

We solve the hydrodynamic equations for the logarithmic density 
$\ln \rho$, the velocity ${\mbox{\boldmath $v$}}$, and the entropy per unit
mass $s$. In terms of these variables, the mass, momentum and
energy conservation equations used to calculate the complete evolution
of a thermally unstable medium are
\begin{equation}
\frac{D \ln\rho}{D t}=-{\mbox{\boldmath $\nabla$}}\cdot{\mbox{\boldmath $v$}},
\label{eq:fluid1}
\end{equation}
\begin{equation}
\frac{D {\mbox{\boldmath $v$}}}{D t} =
-\frac{{\mbox{\boldmath $\nabla$}}P}{\rho}+{\mbox{\boldmath $f$}}({\mbox{\boldmath $r$}},t)
+\mbox{viscous force},
\end{equation}
\begin{equation}
T\frac{D s}{D t} =\Gamma -\rho \Lambda+
\mbox{viscous heating},
\end{equation}
where $D/Dt=\partial/\partial t+
{\mbox{\boldmath $v$}}\cdot{\mbox{\boldmath $\nabla$}}$ is the
advective derivative, 
${\mbox{\boldmath $f$}}$ is a 
forcing function to be specified  that will be used only in
the case of forced simulations (\S \ref{sec:vel_compet}), 
and $\Gamma$ and $\Lambda$ are the
heating and cooling functions, respectively, depending on $\rho$ and $T$.
The equation of state used is that of ideal gases 
$P=(\gamma-1)c_{\rm v}\rho T$,
where $\gamma=5/3$ is the heat
capacity ratio $c_{\rm p}/c_{\rm v}$ for monoatomic gases. 
The specific entropy is related to the density and pressure by 
$s=s_{00}+c_{\rm v}\left[\ln (P/P_{00})-\gamma \ln(\rho/\rho_{00})\right]$, 
where $\rho_{00}$, $P_{00}$ and $s_{00}$ are reference values
for the density, pressure and entropy.

We use simplified forms for the cooling and heating functions,
consisting of a piecewise power-law fit to the ``standard'' equilibrium $P$--$\rho$
curve of Wolfire et al.\ (1995, hereafter WHMTB) and a constant
background heating rate per unit mass. This choice is at an
intermediate level in a hierarchy of increasing complexity in the
treatment of the thermal processes in a gas dynamical
problem, that goes from assuming isothermal (e.g., studies of molecular
clouds by Gammie
\& Ostriker 1996; Mac Low et al.\ 1998; Padoan \& Nordlund 1999; 
Ostriker, Stone \& Gammie 2001) or polytropic flows (e.g.,
V\'azquez-Semadeni, Passot \& Pouquet 1996), to solving the heat
equation using simple fits to the cooling and heating functions
(e.g., Chiang \& Bregman 1988; Rosen \& Bregman 1995;
V\'azquez-Semadeni, Passot \& Pouquet 1995, 1996; Paper I; Burkert \& Lin
2000; Koyama \& Inutsuka 2002; this 
paper), to following the detailed non-equilibrium chemistry and the
cooling rates (e.g., David et al.\ 1988; Murray \& Lin 1989,
1990; Kang et al.\ 1990; Koyama \& Inutsuka 2000). For
the purpose of studying the dynamical development of TI from the
equilibrium state in the neutral ISM,
it is not clear to what extent it is imperative to follow the detailed
non-equilibrium chemistry and the resulting rates (see \S
\ref{sec:comparison} for further discussion). Thus, the present 
work should be taken mainly as an illustration of the highly dynamical
and complex regimes that can already appear within this approximation,
and the caveat of the simplified cooling functions we use should be kept 
in mind.  

Our fit to WHMTB's standard cooling curve has the form
\begin{equation}
\Lambda=C_{i,i+1} T^{\beta_{i,i+1}}  \;\;\;\; {\rm for} \;\;\; T_{i}\leq
T< T_{i+1}.
\end{equation}
As mentioned above, we parameterize the heating function by 
$\Gamma=\Gamma_{0} (\rho/\rho_{0})^{\alpha}$, with
$\alpha=0$, since the dominant heating process (photoelectric heating
from small grains and PAHs) in the density range of interest depends
only weakly on density, roughly as $\rho^{0.2}$ (WHMTB). 

\begin{figure}
\plotone{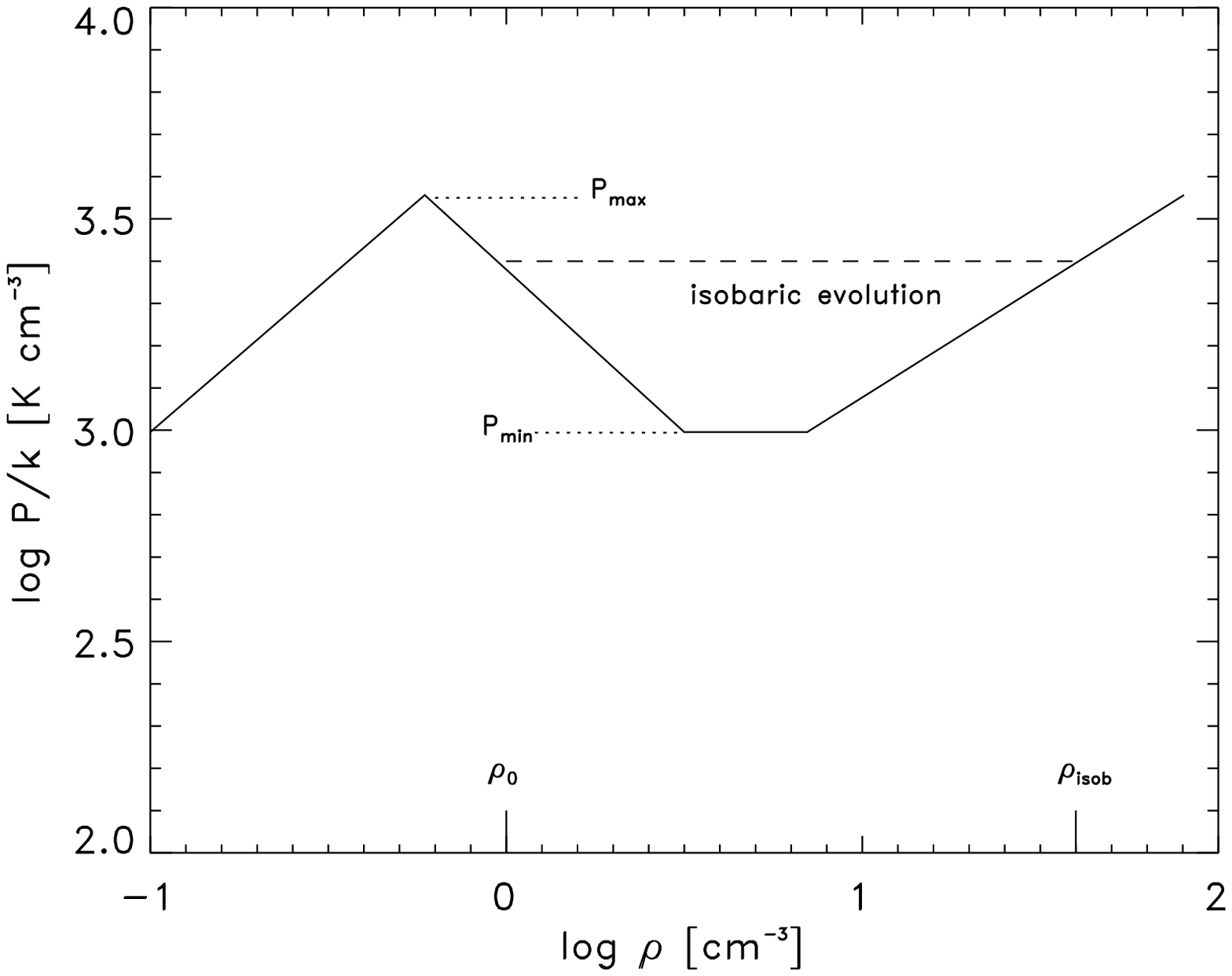}
\caption{Equilibrium curve for the adopted cooling function.
We have labeled as $P_{\rm max}$ and $P_{\rm min}$ the maximum
and minimum pressures in the thermally unstable branch. The unperturbed 
density $\rho_{0}$, and the cloud density resulting from isobaric 
evolution $\rho_{\rm isob}$, are shown.}
\label{cooling}
\end{figure}

We find the following set of 
$\beta$-values and transition temperatures: $\beta_{12}=2.12$,
$\beta_{23}=1.0$, $\beta_{34}=0.56$, $\beta_{45}=3.67$, and
$T_{1}=10$ K, $T_{2}=141$ K, $T_{3}=313$ K, $T_{4}=6102$ K, 
$T_{5}=10^{5}$ K. Note that for $T_{3}< T <T_{4}$
the gas is thermally unstable under the isobaric criterion, and marginally
stable for $T_{2} < T < T_{3}$.
For a constant heating of $\Gamma_0=0.015$ erg s$^{-1}$g$^{-1}$,
the values of the coefficients of the cooling functions in
units of erg s$^{-1}$g$^{-2}$cm$^{3}$K$^{-\beta_{i,i+1}}$ are
$C_{1,2}=3.42 \times 10^{16}$, $C_{2,3}=9.10\times 10^{18}$,
$C_{3,4}=1.11 \times 10^{20}$ and $C_{4,5}=2.00 \times 10^{8}$.
The fitted curve is displayed in Figure~\ref{cooling} 
(c.f. fig.~3 of WHMTB). 

We will denote by $P_{\rm eq}(\rho)$ the pressure corresponding
to thermal equilibrium at density $\rho$, by 
$P_{\rm max}$ the equilibrium pressure
corresponding to the high-temperature boundary of the thermally-unstable 
range, and by $P_{\rm min}$ the low-pressure plateau in the thermal
equilibrium curve (see Fig.~\ref{cooling}).
As ``standard'' unperturbed initial conditions we take $\rho_{0}=1$
cm$^{-3}$, representative of the mean density in the ISM, and $T=2400$
K, the thermal equilibrium temperature corresponding to $\rho_0$. This
state falls in the unstable range. Other workers have chosen to apply
nonlinear perturbations 
(strong compressions or shocks; e.g., Hennebelle \& P\'erault 1999; 
Koyama \& Inutsuka 2000, 2002) to linearly {\it stable} gas in order to
determine whether TI can be triggered in such regimes by strong
perturbations. However, since in this paper we are interested in
the conditions necessary for the development of TI to be highly dynamic
and in the
competition between modes, we prefer to take the initial conditions
described above, which, we remark, are close to the mean density of
the ISM (see, e.g., Ferri\`ere 2001).

We use a sixth-order finite difference scheme for the spatial
derivatives on a one-dimensional, non-uniform Cartesian grid, and a third-order
time stepping scheme, as described in S\'anchez-Salcedo
\& Brandenburg (2001). The code includes a shock-capturing viscosity
proportional to $-{\mbox{\boldmath $\nabla$}}\cdot {\mbox{\boldmath $v$}}$
wherever ${\mbox{\boldmath $\nabla$}}\cdot {\mbox{\boldmath $v$}}<0$, 
and zero otherwise. Shocks are resolved over $3$--$4$ zones. To avoid
numerical instability due to amplitude errors at high-wavenumber
modes of density and entropy, diffusion terms in the continuity
and entropy equations have been introduced. Convergence tests (\S
\ref{sec:convergence}) have
been performed to ensure that mass and thermal diffusion do not affect the
growth rate of forming density peaks. In
particular, the effective mass diffusion scale implies that 
below scales $\sim 10$ pixels the growth rate starts to
decrease monotonically (see also Paper III).

Most of the simulations presented here
assume a plane-parallel geometry. This is motivated by
the need to achieve very high resolutions, but is justified by
evidence that suggests the presence of sheet-like and filamentary morphologies
in the cold neutral medium (e.g. Kim et al.~1998). Although the origin
of this morphology may not be caused by TI, we consider this geometry to
be more appropriate than, for example, a spherical one for studying the
development of TI in the neutral medium of the Galaxy. We discuss this
issue further in \S \ref{sec:lim_1d}.

The non-uniform grid allows for enough resolution as to resolve 
the structure of the cloud and the accretion fronts. 
It also allows us
to move the boundaries far away, so no perturbation has time
to reach them. For simulations of forced flows, we
use a uniform mesh with periodic boundary conditions.

\subsection{Resolution requirements} \label{sec:resol}

One condition on the the minimum resolution necessary in the
simulations is obtained by requiring that the final condensations be
well resolved.
Consider a domain of size $l$ per dimension covered with a uniform mesh 
containing $n_x$ grid points, in the $x$-direction, 
separated by a distance $\delta x$, and
an initial density perturbation with a characteristic size $l_{0}=\psi\, l$. 
In 1D, the final size of the condensation, $l_{1}$, will be  
$l_{1}\approx l_{0}\rho_{0}/\rho_{1}= \psi\, l \rho_{0}/\rho_{1}$, 
where $\rho_0$ and $\rho_1$ are respectively the mean and final densities.
If we require a minimum number of zones, $n$, inside $l_{1}$,
the total number of grid points must be larger than 
$n\rho_{1}/(\psi\rho_{0})$. For $\rho_{1}/\rho_{0}=50$, 
$n=10$, and $\psi=0.5$, we need $n_x>1000$.
If this condition is not met, our tests indicate that a stationary
regime is reached in which mass diffusion balances accretion, leading
to a very unrealistic situation.

If one wants to resolve the out-of-equilibrium thermal structure 
within accretion fronts,
a second condition must be satisfied, namely that $\delta x$
must be smaller than the cooling length. 
The cooling length is given by $l_{\rm c}=\tau_{\rm cool}v_{\rm fr}$, where
$\tau_{\rm cool}$ is the cooling time and $v_{\rm fr}$ the front velocity. 
The condition reads
$n_x>l/(\tau_{\rm cool}v_{\rm fr})$. This is a highly variable
condition, as both $\tau_{\rm cool}$ and $v_{\rm fr}$ 
decrease with
increasing density, so the condition is most stringent at high
densities. For  $\tau_{\rm cool}\approx 0.2$ Myr 
and $v_{\rm fr}=0.1$ km s$^{-1}$ we need 
$n_x>10^{4}$ for $l=100$ pc and $n_x>10^{5}$ for $l=1$ kpc.
The need for huge resolution can be alleviated by using a
non-uniform grid, with a maximum density of points where the
condensation is expected to form.

An alternative approach is to reduce the heating and cooling
coefficients by some factor. 
This modification of the cooling rates is justified only 
if $\eta \ll 1$ at all scales of interest, and at all times, since, in
effect, this is equivalent to reducing the simulation box size.
However, in regions where the gas is evacuated, the density decreases and
the cooling times increases. Hence, it is not true that the condition
$\eta\ll 1$ holds at all times in the warm diffuse
gas, and we cannot reduce the heating and cooling
coefficients by the same factor everywhere, but instead we would have to 
use a density-dependent 
function which reduces the cooling rates only at very high densities. 
For these reasons, we prefer to stick to non-uniform grids
in simulations of single condensations, and to very-high-resolution
uniform grids in multi-fluctuation simulations. Note, however, that
these high resolution requirements arise only in order to resolve such
small-scale structures as the cloud formed or the thermal structure
within radiative shocks. Nonetheless, much lower resolutions still
suffice to describe the global dynamics. 

\section{Time scales and general features} \label{sec:quali}

In this Section, we describe some basic concepts useful for
understanding the remainder of the paper, including the characteristic 
time scales involved, as well as some features of the growth of
fluctuations. We also define the relevant time scales explicitly in the
cases of density and velocity fluctuations.

\subsection{Modes of the instability and the dynamical evolution of
fluctuations} \label{sec:modes_growth}

As mentioned in the Introduction, in this paper we are concerned
with cooling functions which imply instability only for the
isobaric mode. In the presence of
constant heating ($\alpha=0$), this is satisfied for $\beta <1$, in which
$\beta$ is the exponent of the temperature in the cooling function (see \S
\ref{sec:model}). For our fit to the standard cooling function of
WHMTB, this occurs in the temperature range $300$ K
$\lesssim T \lesssim 6100$ K, where $\beta =0.52$, i.e., between the so
called ``cold'' and 
``warm'' phases of the ISM. In the presence of constant heating, the
isochoric mode of TI is unstable for $\beta  
<0$, and thus for temperatures between $\sim 10^5$ and $\sim 10^6$ K for
classical cooling functions (e.g., Raymond, Cox \& Smith 1976). Such
temperatures are not reached by our simulations, since we do not
include supernova heating. Finally, it can be easily shown that in our
case the adiabatic criterion requires $\beta < 1/(1-\gamma)=-3/2$ for
instability, and thus our simulations are stable under the latter
two criteria in the vicinity of thermal equilibrium. Note that, however,  
far from thermal equilibrium, the system may cool freely, and then the
isochoric mode may also become unstable, as in the cooling case (i.e.,
in the absence of background heating), in which the instability criteria
for the 
isobaric and isochoric modes are, respectively, $\beta < 2$ and $\beta < 
1$.

\subsection{Evolution of isobaric density perturbations} \label{sec:entr_pert}

Here we consider the growth of initial states with density perturbations
of a given physical size, no fluid motions and constant
pressure. Isobaric density perturbations constitute a simple case of entropy
perturbations, as they imply a perturbation of the ratio
$P/\rho^\gamma$, and can be produced in practice by either locally
varying the cooling or heating rates, or by velocity fluctuations well
into the subsonic regime (see \S\S \ref{sec:vel_fluc} and
\ref{sec:vel_compet}). 

Let us consider, for simplicity, the condensation process
of a perturbation $\rho_{1}=\rho-\rho_{0} >0$.   
For a linear perturbation of wavelength $\lambda$ there are two relevant
timescales at $t=0$: the sound crossing time 
$\tau_{\rm s}(0)\equiv \lambda/c_{\rm
s}(0)$, where $c_{\rm s}(0)$ is the adiabatic sound speed, and the cooling time
$\tau_{\rm cool}(0)\equiv c_{\rm v}/\left|\left(\partial\rho
\Lambda/\partial T\right)_P\right|$, which is of the order
of $\approx e_{0}/\rho_{0}\Lambda$, 
where $e_{0}$ is the initial internal energy per unit mass. 
The initial time scale ratio is then
$\eta_{0}\equiv \tau_{\rm cool}/\tau_{\rm s}$. 

It is well
known (Field 1965; see also Shu 1992; Balbus 1995; Meerson 1996;
Burkert \& Lin 2000) that
entropy perturbations with $\eta>1$ evolve almost quasi-statically,
maintaining quasi-isobaricity throughout their evolution. 
On the other hand, entropy perturbations with $\eta <1$  cool
before they can equate pressures and therefore
develop large pressure gradients, which then cause them to evolve
highly dynamically even if only the isobaric instability criterion is
satisfied. It is also well known that perturbations with $\eta > 1$ 
grow on time scales $\sim \tau_{\rm cool}$, because in this case the
perturbation can only evolve as fast as it cools, and thus the growth
rate saturates at $\sim 1/\tau_{\rm cool}$ in the limit of small
perturbation scales. Instead, when only the isobaric criterion is
satisfied, perturbations with $\eta < 1$ grow on a time $\sim \tau_s$,
because they can only evolve as fast as the pressure gradient pushes the 
gas surrounding the condensation to equate pressures (see Field 1965
and Meerson 1996 for more details). This implies that the growth rate
approaches zero in the limit of long wavelengths. In order for
finite growth rates to exist in this limit, it is necessary to also
satisfy the isochoric criterion.

If the evolution were strictly isobaric, a condensation (cloud) would
end up acquiring
the density denoted $\rho_{\rm isob}$ in Figure~\ref{cooling}.
Adopting the cooling function described in \S \ref{sec:num_meth} and
a mean 
interstellar density of $1$ cm$^{-3}$ ($T_{\rm eq}\simeq 2400$ K),
$\rho_{\rm isob}=40$ cm$^{-3}$. Furthermore,
the condition $\tau_{\rm s} \sim \tau_{\rm cool}$ occurs at a
size scale $\lambda\simeq 10$ pc (hereafter denoted by $l_{\rm eq}$).
Therefore, isobaric evolution cannot
occur at scales larger than a few parsecs. 
Let us thus consider a perturbation (of large enough size scale)
such that $\eta \ll 1$. Since $\tau_{\rm cool}
\propto \rho^{-1/\beta}$ and $\tau_{\rm s}$ cannot decrease faster than
$\rho^{-1/D}$ during the condensation
process, where $D$ is the dimension of the contraction, we 
conclude that $\eta \ll 1$ in the core of the cloud (i.e.~at
the vicinity of $x\simeq 0$) at all subsequent
times, even as it shrinks. 
This is in sharp contrast with the fact that smaller scales have
larger values of $\eta$, under the mean conditions of the unperturbed
medium.

Relatively high local Mach numbers can be attained in the condensation
process under conditions with $\eta_{0}\ll 1$. An estimate of the
maximum velocities that can be reached due to pure TI 
can be made from Bernoulli-like arguments as
$v_{\rm inst}^{2}\approx 2(P_{\rm max}-P_{\rm min})/\rho_{0}$,
leading to fluid velocities $v_{\rm inst}\sim 4$ km s$^{-1}$,
for $\rho_{0}=1$ cm$^{-3}$, $P_{\rm max}/k=3000$ K cm$^{-3}$ and
$P_{\rm min}/k=1000$ K cm$^{-3}$. This velocity is clearly
supersonic in a medium at the temperature corresponding
to $P_{\rm min}$, at which $c_{\rm s}\approx 2$ km s$^{-1}$. 
With that typical velocity, a region of size $L$ ($L_{0}$
at time $t=0$) condenses on a characteristic timescale
\begin{equation}
\tau_{\rm inst}=\frac{L_{0}}{v_{\rm inst}}= 
1.2 \left(\frac{L_{0}}{10 \;{\rm pc}}\right) 10^{6}\;\;\;{\rm yr}.
\label{eq:tau_inst}
\end{equation}

The characteristic time given by Eq.\ (\ref{eq:tau_inst}) can be
compared to the free-fall time of a uniform
cloud due to self-gravity,
\begin{equation}
\tau_{\rm ff}=\left(\frac{3\pi}{32 G\rho}\right)^{1/2}=
4.4 \left(\frac{\rho}{1 \; {\rm cm}^{-3}}\right)^{-1/2} 10^{7}\;\;\;
{\rm yr}.
\end{equation}
Therefore, self-gravity must be included either at scales
larger than $\sim 0.4$ kpc 
or at late stages due to the density enhancement. In a plane-parallel
condensation $\tau_{\rm ff}(t)\propto L^{1/2}$, so that
the ratio $\tau_{\rm ff}/\tau_{\rm inst}$ increases 
during the condensation process. Thus, for the fiducial values we have
used, the flow may reach the thermally 
stable regime and re-approach pressure equilibrium, leaving only a
slow-accretion regime (see \S \ref{sec:ev_density}), without self-gravity 
ever becoming dynamically important. 
Consequently, overshooting of density will be unable to
prompt gravitational instabilities in 1D experiments, but it may be
important in more than 1D (e.g., Schwarz et al.\ 1972; Kang et al.\ 2000).

\subsection{Fluctuations in the velocity field} \label{sec:vel_fluc}

If, in contrast with the cases described above, an initial state of
constant density is perturbed with velocity
fluctuations (of amplitude $v_{0}$), 
the evolution may be quite different. The
generalization of the instability analysis to a non-static initial state 
was performed by Hunter (1970, 1971). Velocity
perturbations as triggers of TI in initially stable media have been
investigated by a number of
authors (e.g., Murray \& Lin 1989; Kang et al.\ 1990; Hennebelle \&
P\'erault 1999, 2000; Koyama \& Inutsuka 2000), but here we are
interested in the opposite 
effect, i.e., whether perturbations induced by velocity fluctuations in
initially unstable media can actually be stable, as first suggested, but
not shown, in Paper I. A thorough
discussion has been presented in Paper III, and here we just repeat the 
main points.

In the presence of velocity fluctuations, a new characteristic time
scale appears in the system, namely the bulk-velocity crossing
time, $\tu =\lambda/u$, where $u$ is the characteristic velocity
difference across scale $\lambda$. Thus, the dynamical
time $\tau_{\rm dyn}$ to use in $\eta$ should be 
chosen as the minimum of the sound and the bulk velocity crossing times. 

Compressive motions in the absence of radiative cooling cause adiabatic
perturbations in which
the density and pressure fluctuations have the same sign, because of the $PdV$ 
work done on the affected fluid parcel. In the presence of cooling, the
actual behavior depends on (the redefined) $\eta$. If the cooling time
is much longer than the characteristic time of the compression (i.e., if 
$\eta \gg 1$), the cooling is negligible, and the compressed parcel behaves
as a quasi-adiabatic sound wave. This case is known to be unstable only if the
adiabatic criterion is satisfied, which for our piecewise power-law cooling
function requires $\beta < 1/(1-\gamma) = -3/2$. In the opposite case
($\eta \ll 1$), the $PdV$ 
heating is negligible in comparison with the cooling, and the pressure 
is determined by the thermal equilibrium condition. In this limit, the
velocity perturbations are unstable when the isobaric condition is
satisfied, because they behave as a condensation mode (see, e.g., Shu 1992;
Paper III). Thus, velocity fluctuations strong enough that they dominate
$\eta$ are linearly stable at $\eta >1$ (small scales) but unstable at
$\eta < 1$ (large scales) in the atomic ISM. Moreover, when $\tu<
\tau_{\rm s}$, the 
instability of large-scale perturbations manifests itself in a null
resistance of the flow to turbulent compressions rather than in
spontaneous growth of the perturbations, because the bulk velocity is
larger than the sound speed, and so the 
driver of the density growth is the turbulent velocity compression
rather than the thermal pressure gradient.

It is worth remarking that velocity fluctuations are 
a natural way of producing the density fluctuations that constitute the
initial conditions for the subsequent development of TI,
because the gas obeys to the continuity equation (mass
conservation), and thus the production of a density enhancement requires a 
converging velocity field. 

The case of strongly nonlinear compressions probably depends on the
closeness of $\eta$ to unity. If $\eta \gg 1$, then even very strong
compressions behave adiabatically, and are not able to induce very large 
density enhancements. If $\eta$ is still larger than
unity at the higher density, triggering the isobaric mode will be
extremely difficult. On the other hand, if $\eta$ is
not much larger than unity initially, then the density increase induced
by the compression may raise the cooling rate to large enough values
that $\eta$ becomes smaller than unity, and triggering of the isobaric
instability ensues.

\section{Numerical results} \label{sec:front_dyn}

\subsection{Evolution of a single density perturbation and final
stages of TI} \label{sec:ev_density}

In this section we attempt to quantify the dynamical nature of the
condensation process in the context of the
formation of interstellar clouds by TI. In particular, we seek to
determine the physical perturbation sizes at which the condensation
develops supersonic velocities, the times needed for these velocities to 
subside, and the state of the leftover diffuse gas, which, as we shall
see, contains significant fractions of mass still in the unstable density 
range.

Let us consider first how the nonlinear evolution of a
Gaussian density perturbation depends on
$\eta_{0}$. To this end, we numerically follow the evolution of
perturbed density fields of the form 
\begin{equation}
\rho_{0}(x)=1+A_{0}\exp\left(-\frac{{x}^{2}}{2\sigma^{2}}\right),
\end{equation}
where the density is in units of cm$^{-3}$.
We take $A_{0}=0.2$ and consider three values of 
$\sigma$ such that the FWHM of the Gaussian
profile is $3$, $15$ and $75$ pc.
We refer to the corresponding simulations as DEN3, DEN15 and DEN75.
According to our discussion in \S \ref{sec:entr_pert}, it is expected that for
model DEN3, the evolution will be closer to isobaric and quasi-static
($\eta_{0} \approx 3.7$, assuming $\tau_{\rm s}={\rm FWHM}/2 c_{\rm s}$),
whereas dynamical compression regimes are expected for DEN15 
($\eta_{0}\approx 0.75$) and DEN75 ($\eta_{0}\approx 0.15$). 
All these simulations start with zero velocity.

\begin{figure}
\plotone{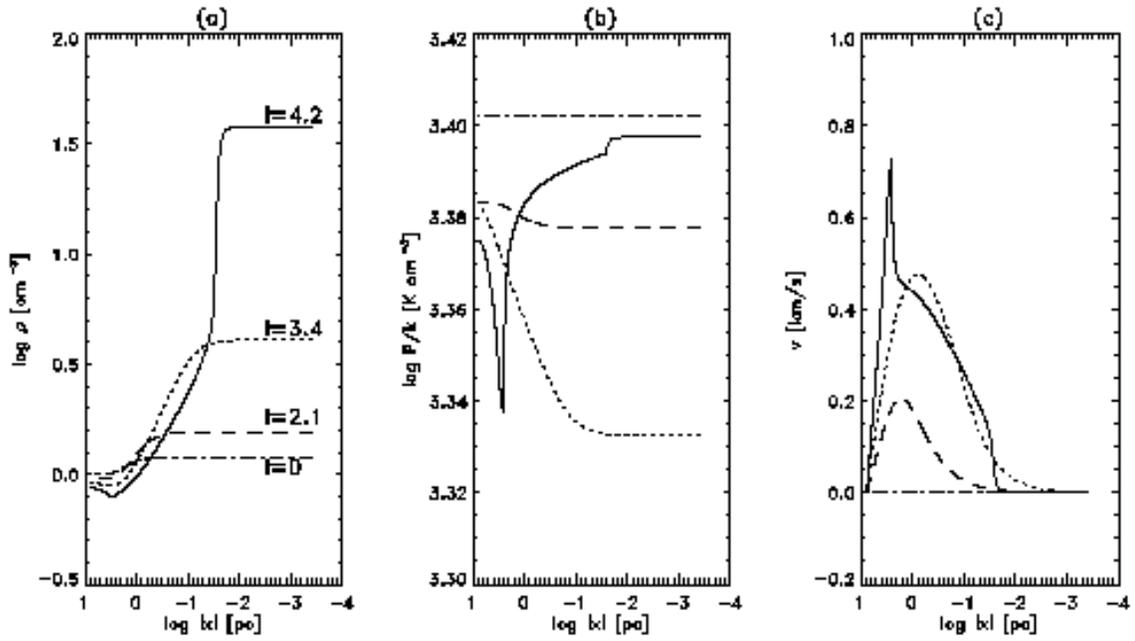}
\caption{
Time development of run DEN3. The density,
pressure and velocity profiles at different times are plotted in panels 
(a), (b) and (c), respectively. 
In all three panels the initial values are drawn
with dash-dotted lines, at $t=2.1$ Myr with dashed lines, at
$t=3.4$ Myr dotted lines and at $t=4.2$ Myr with solid lines.
}
\label{den3a}
\end{figure}

Model DEN3 starts with constant
pressure. The initial pressure is $P/k=2500$ K cm$^{-3}$, which
equals the equilibrium pressure $P_{\rm eq}$ at $1$
cm$^{-3}$. Figure~\ref{den3a} shows the density, velocity and pressure
profiles at four different temporal snapshots. The 
maximum of the velocity, $v_{\rm max}(t)$, increases continuously 
until $t\simeq 3.8$ Myr, at which time $v_{\rm max}\approx 0.86$ km/s 
(Mach number $0.18$). 
As expected, although the pressure is not perfectly constant
(see Fig.~\ref{den3a}b), the evolution can be considered as
quasi-isobaric, as can be seen in Figure~\ref{den3b}. Indeed, the
maximum pressure variation is $\left|\delta P\right|/P_{0}\approx 0.2$.
As a consequence, there is no appreciable density
overshoot beyond $\rho_{\rm isob}$ (\S \ref{sec:entr_pert}), and
the final pressure inside the cloud recovers a value 
very close to the ambient pressure. 

By time $t\approx 4$ Myr, a nearly stationary ``slow-growth'' stage is
reached, in which most of the mass is in the central condensation at
nearly zero velocity, with low density material falling in from the
surrounding medium, but decelerating smoothly as it enters the cloud,
without the formation of shock. The time for reaching this stage 
increases significantly when the initial perturbation amplitude
is decreased ($\sim 5.5$ Myr for $A_0=0.1$).  
In order to quantify the amount of mass within the condensation, 
we integrate the density over the length of the region where $\rho \ge
10 \rho_0$, leading to a column density\footnote{Note
that the units of column 
density are M$_{\odot}$ pc$^{-2}$ even though the simulations are
one-dimensional, because the physical units of the density are
g cm$^{-3}$ anyway.} $N\sim 0.045$ M$_{\odot}$ pc$^{-2}$ 
at $t=4.2$ Myr. This value is 67\% of the column density 
initially within $|x|<\sigma$.

It is important to remark that in the ``final'' slow-growth state mentioned 
above, the leftover diffuse gas is far from static, with velocities of up to
0.7 km s$^{-1}$. Although the accretion rate onto the condensation is
low because of the low density of the accreting gas, the amount of mass
contained in this regime is still large, $\sim 8$ times larger than the
amount of mass in the condensation. This result is in part an artifact
of the fact that we considered a single perturbation, which by itself
does not consume a large amount of mass by the time it reaches
$\rho_{\rm isob}$. Smaller final fractions of mass still in the unstable 
range are expected in the case of multiple perturbations, but even in
this case, much longer times are needed to evacuate the unstable range
(\S \ref{sec:scale_compet}). Smaller unstable fractions are also expected
if the background medium is already in the stable diffuse phase and the
condensation is triggered by a strong compression (we thank P.\
Hennebelle for pointing this out).

\begin{figure}
\plotone{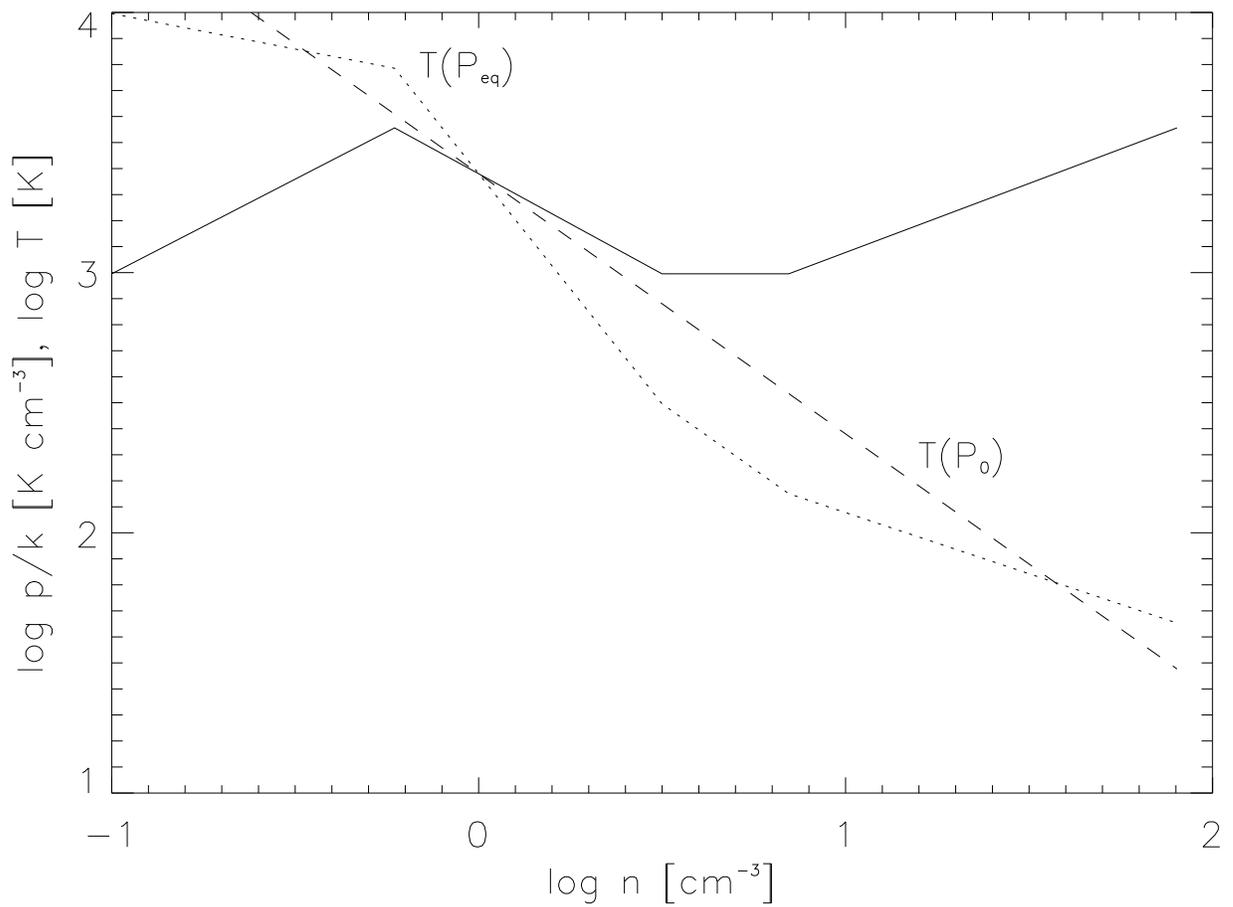}
\caption{Comparison of the temperatures corresponding to intermediate
densities under the thermal-equilibrium (dotted line) and isobaric
(dashed line) regimes. The leftover accreting gas at the end of the
condensation process flows along the nearly isobaric regime.
}
\label{fig:isob_vs_eq}
\end{figure}

\begin{figure}
\plotone{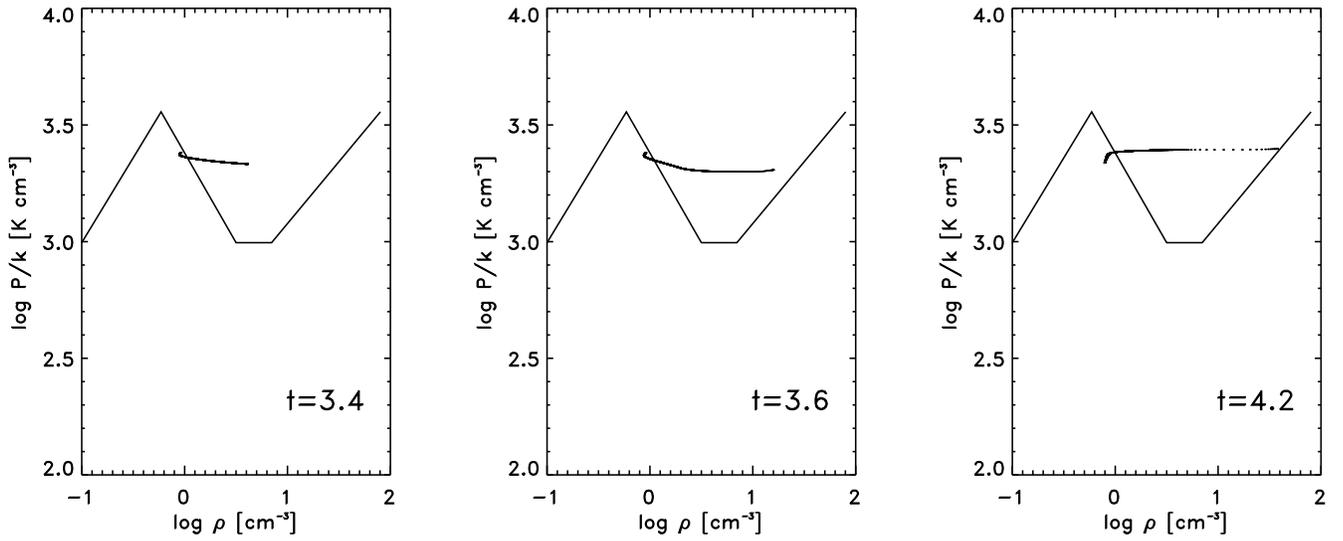}
\caption{Pixel by pixel plot of $\log(P/k)$--$\log(\rho)$ for run
DEN3 at different times.  
For reference, the equilibrium $\log(P/k)$ curve (see Fig.~\ref{cooling})
is also plotted. Time is shown at each panel
in units of $1$ Myr.  Note that two of the diagrams were taken at the same time
than the snapshots of Fig.~\ref{den3a}. 
}
\label{den3b}
\end{figure}

Another point worth remarking is that the accreting gas evolves in a 
nearly isobaric regime in response to the instability, but such that the
small existing pressure
variations have the same sign as the density variations
(Fig. \ref{den3a}), and thus this gas has no further tendency to
fragment, even though its density lies in the ``unstable''
range. Moreover, this implies that this density range does not precisely 
correspond to the unstable temperature range, as shown in Fig.\
\ref{fig:isob_vs_eq}.

\begin{figure}
\plotone{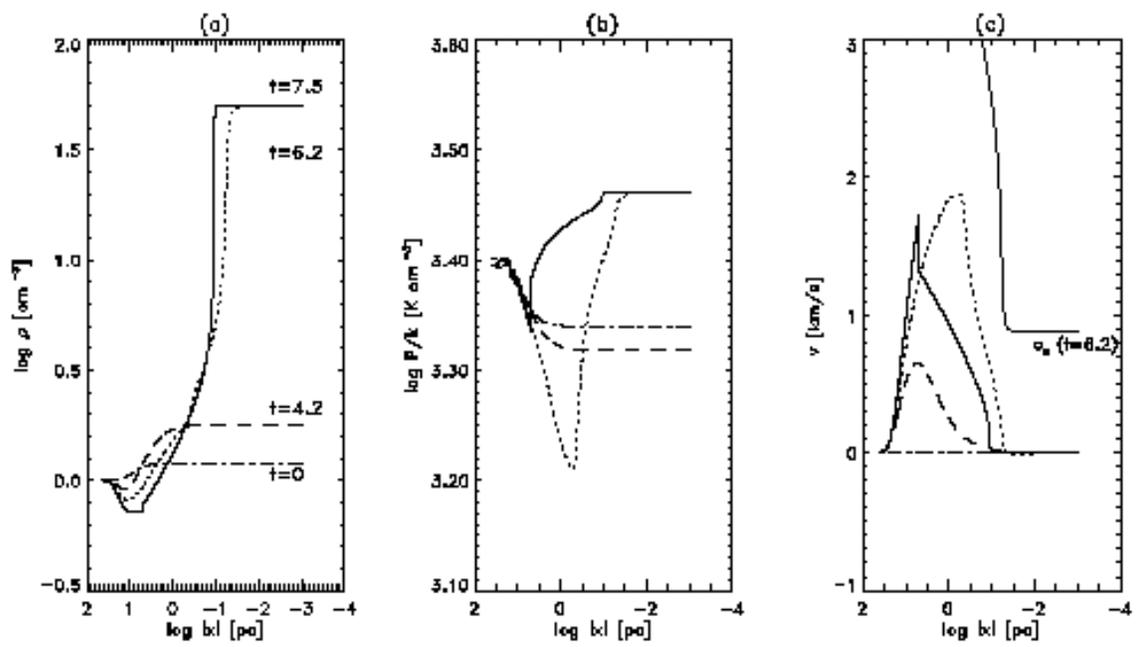}
\caption{Same as Fig.~\ref{den3a} but for run DEN15. In panel
(c) the curve labeled with $c_{\rm s}$ represents the adiabatic sound
speed at $6.2$ Myr.}
\label{den15a}
\end{figure}

\begin{figure}
\plotone{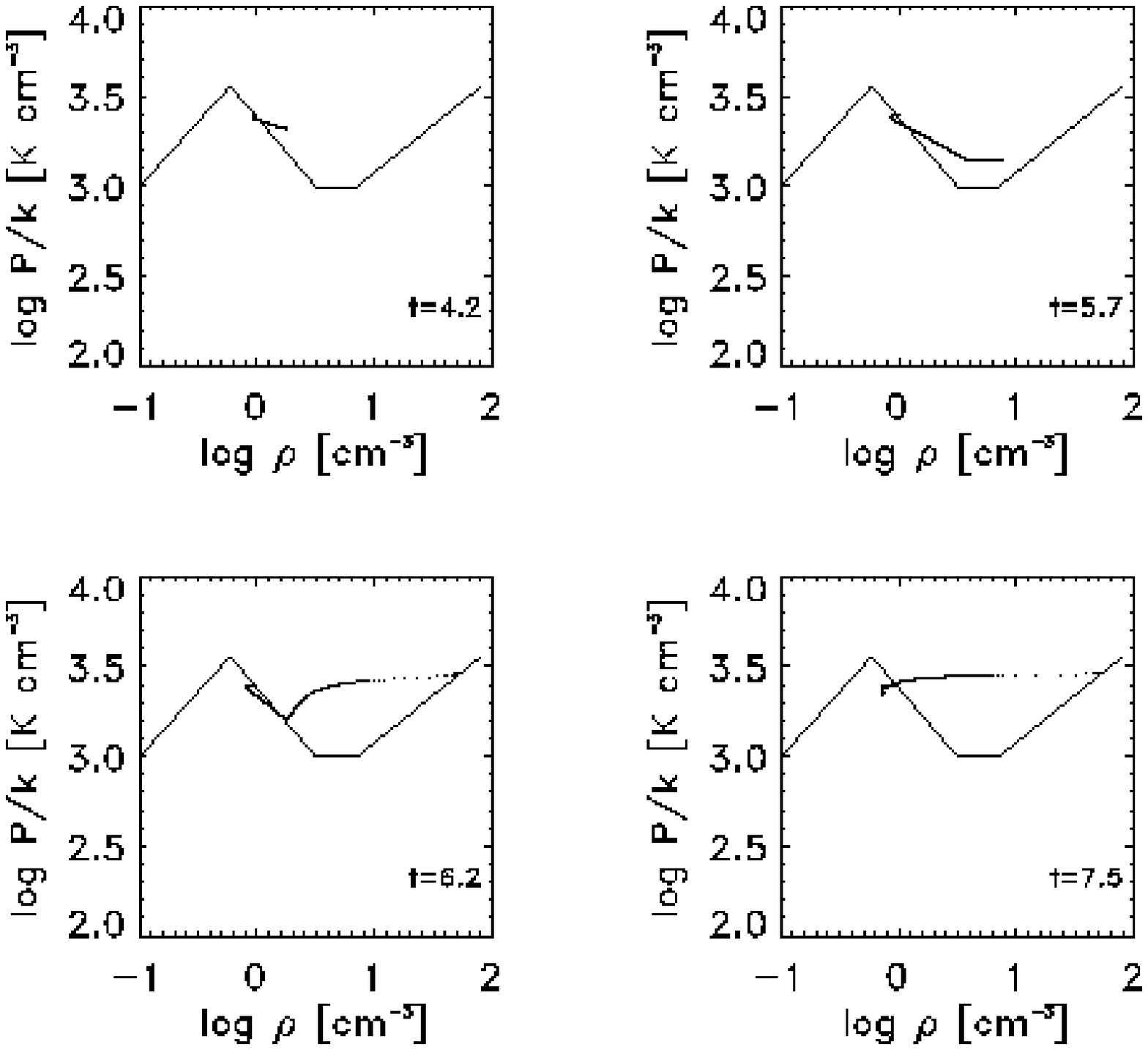}
\caption{Same as Fig.~\ref{den3b} but for run DEN15.
Note that three of the diagrams were taken at the same time
than the snapshots of Fig.~\ref{den15a}.
}
\label{den15b}
\end{figure}

Let us now consider the case of  models DEN15 and DEN75. In these, we 
set up the initial pressure to the corresponding 
$P_{\rm eq}$ values, $P_{0}=P_{\rm eq}\left(\rho_{0}(x)\right)$, 
which is reasonable for models with low $\eta$'s, i.e. short cooling times. 
Figure~\ref{den15a} displays four
snapshots of DEN15. We can distinguish three evolutionary stages:
(1) An initial compression stage in which all the parcels in the flow
are in the thermally unstable range, the fluid is continuously
accelerated, and the pressure gradient becomes increasingly steeper,
eventually generating transonic infall  motions. (2) A
repressurization, or ``crushing'', stage, starting when
the core of the cloud enters the thermally stable temperature range
and the pressure 
starts to increase in the center of the cloud, while the outside
material is still unstable and continues to accelerate inwards,
generating a shock at the edge of the cloud at the time of maximum
compression. Note that this 
is a radiative shock occurring over distances of the order of the cooling
length, and so it is well resolved by the numerical grid.
(3) A slow, subsonic accretion stage, which slowly
fades away. During the 
initial compression stage the system is closer to thermal equilibrium
than in DEN3 (see Fig.~\ref{den15b}). 
The maximum Mach number achieved in this
run was $0.6$, and occurred at $6.1$ Myr.
It is necessary to start with a perturbation of FWHM of $40$ pc for 
the infall velocity to be transonic.

The shock wave that emanates from the cloud boundary is worth discussing.
This shock is rather weak,
as can be seen from the pressure jump in Figure~\ref{den15a}b,
but has important effects in the warm medium. As
shown in the $P$--$\rho$ diagram (Fig.~\ref{den15b}), 
as the shock propagates outwards, it
heats the inflowing gas, bringing it closer to an isobaric regime,
similar to the one of run DEN3, while
bringing it out of thermal equilibrium. 
In fact, the relative pressure variation $\delta P/P_{0}$ from $x=0$ to
the position just behind the shock is only $\sim 0.2$ in
the slow accretion stage, whereas the temperature varies from $50$ to
$3000$ K over this same interval. The column density of the cloud
is $0.26$ M$_{\odot}$ pc$^{-2}$, about $76\%$ of the initial
column density within $|x|<\sigma$ ($0.335$ M$_{\odot}$ pc$^{-2}$).
Since the ram pressure in the boundary of the cloud is larger than
in model DEN3, the cloud density is slightly larger too.

\begin{figure}
\plotone{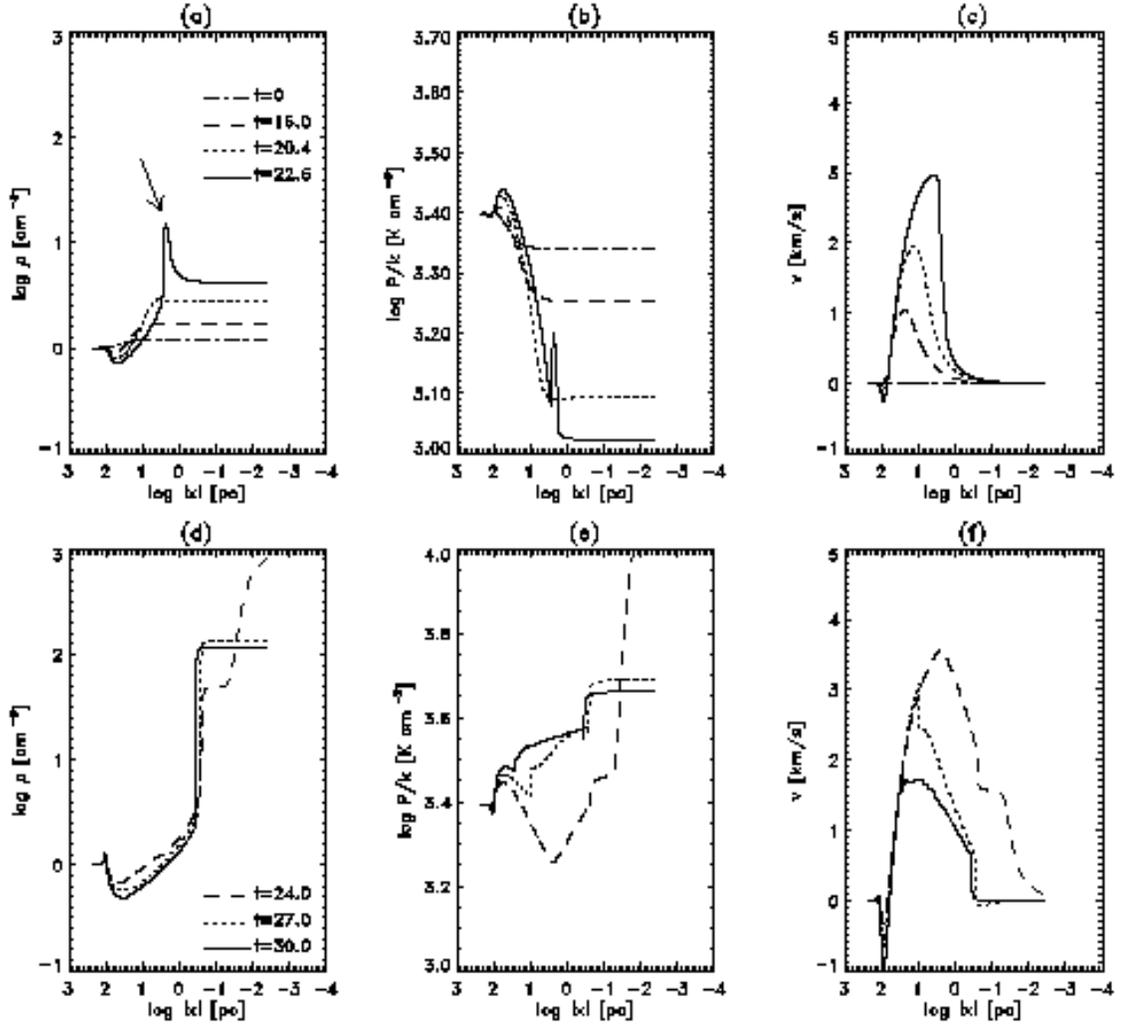}
\caption{
Time development of run DEN75. The density,
pressure and velocity profiles at different times are plotted in panels 
(a) and (d), (b) and (e), and (c) and (f), respectively. The arrow
in panel (a) marks the location of a peak in density originating from
the small-scale components of the initial gaussian fluctuation (see text). 
}
\label{den75a}
\end{figure}

\begin{figure}
\plotone{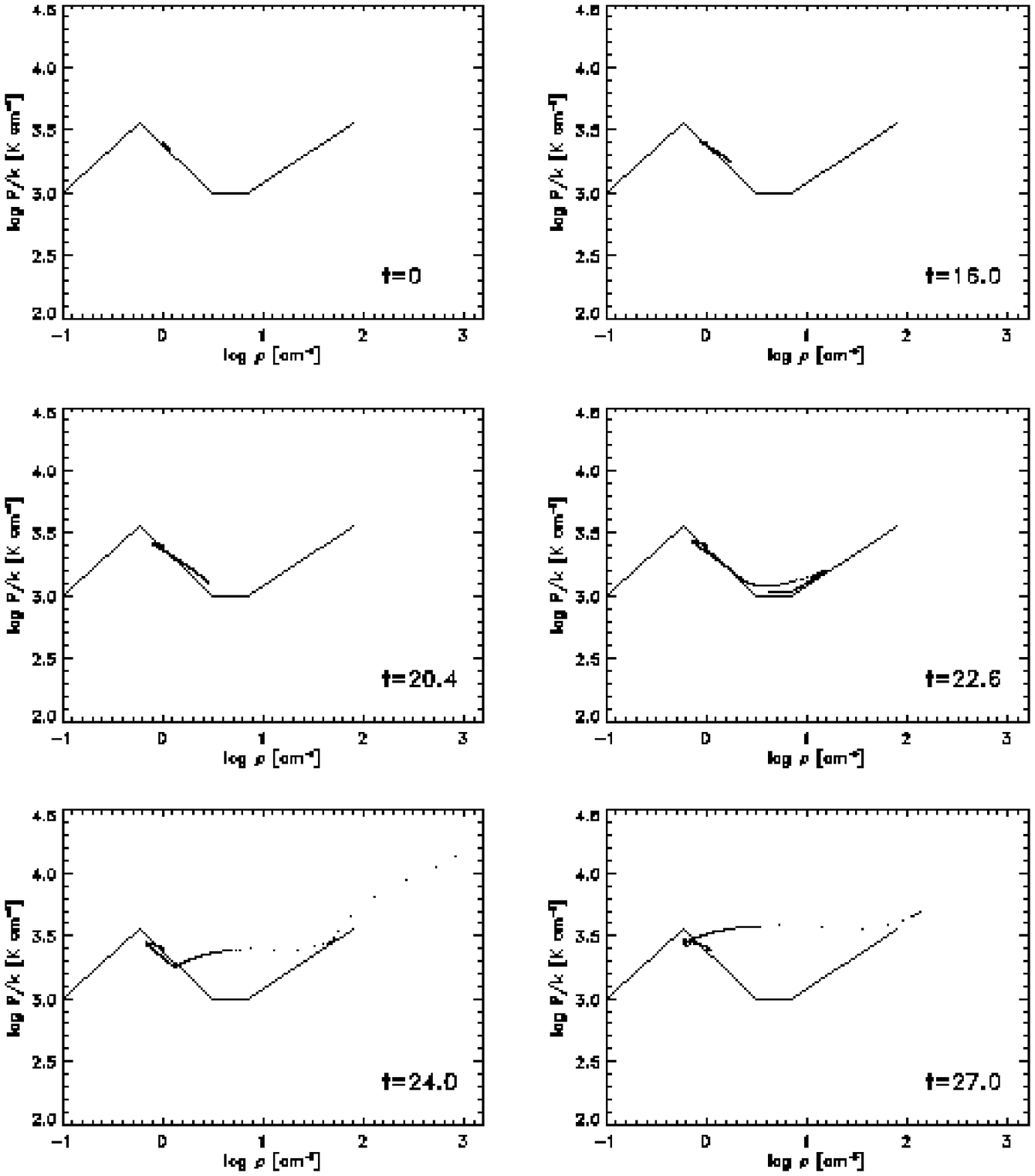}
\caption{Same as Fig. \ref{den3b} but for run DEN75.
Note that the diagrams were taken at the same time
than the snapshots of Fig.~\ref{den75a}.
}
\label{den75b}
\end{figure}

We now consider run DEN75, for which $\eta_{0} \approx
0.15$. Figure~\ref{den75a} 
shows the evolution of density, pressure and velocity. The three
evolutionary stages discussed in model DEN15 are also present in DEN75.
In this run, the gas behaves precisely along the thermal equilibrium
$P$-$\rho$ curve in the initial collapsing stage ($t\leq 21$ Myr), as we
can see in Figure~\ref{den75b}. In this situation,
the gas acquires the maximum possible values of the infall velocity
imposed by the development of pure TI. In fact, at the time 
when the cloud begins to reach the thermally stable phase ($T\approx 140$ K),
the maximum velocity is $3.0$ km s$^{-1}$, and the 
maximum local Mach number is $1.2$. The maximum of the velocity
continues increasing up to $3.6$ km s$^{-1}$, agreeing with our
estimate for $v_{\rm inst}$ in \S \ref{sec:entr_pert} to within 10\%.
Consequently, the supersonic crushing
process is more violent than in DEN15, 
causing a large density overshoot, of roughly $30 \times \rho_{\rm
isob}$. Note that the actual value of the density peak in these
simulations is limited by shock viscosity and mass diffusion, although
this value is of little interest in
1D contractions because, as mentioned in \S \ref{sec:entr_pert}, it is not 
viable for triggering gravitational instabilities, except, perhaps, if
the densities are high enough to trigger a non-equilibrium instability
in the cold gas due to molecular cooling (Koyama \& Inutsuka 2000),
which we do not include here. The density of the cloud will subsequently
decrease slowly with time, but it will remain significantly
larger than $\rho_{\rm isob}$ (see below).

During the compression phase, the small-scale Fourier components of
the Gaussian perturbation blow up due to their larger growth rates,
forming two (symmetric) 
prominent density peaks, which ultimately merge with each other 
(Fig.\ \ref{den75a}; note that only one peak is seen, since only one side of
the simulation is shown).
Their formation depends on details of the initial
conditions, being nonexistent in the presence of a small initial
convergent velocity field. The overall evolution of a simulation (DEN75b
hereafter) with FWHM=75 pc,
$A_{0}=0.7$ and initial velocity field 
$v_{0}\propto -\nabla P_{0}/\rho_{0}$, with initial maximum
Mach number of $0.4$, does not present remarkable differences
except for the presence of stronger accretion.

\begin{figure}
\plotone{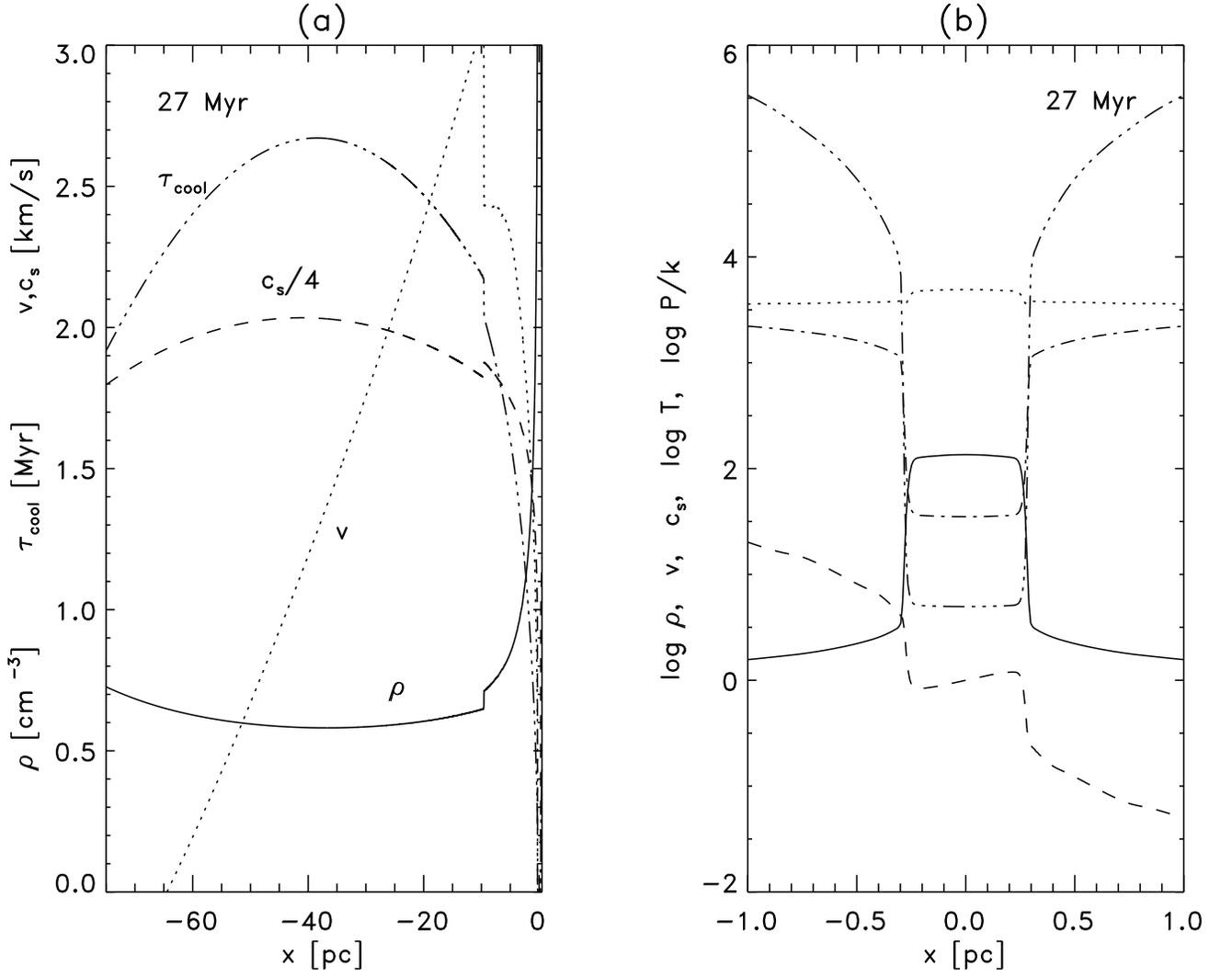}
\caption{
Profiles of the density, sound speed (divided by 4), velocity,
and cooling time profiles, between $0$ and $-75$ pc at $t=27$ Myr for 
run DEN75 (panel a).  Panel (b) shows a 
zoom of the central parsec of this run.
The density ({\it solid line}) is in cm$^{-3}$, the velocity  
({\it dashed line}) and the adiabatic sound speed ({\it triple-dot dashed 
line}) are in km/s, $P/k$ ({\it dotted line}) is in K cm$^{-3}$,
and the temperature ({\it dot-dashed line}) is in K.
Note that the simulation is symmetric with respect the origin.} 
\label{den75c}
\end{figure}

As with run DEN15, at later times, the shock wave propagates outwards
from the cloud boundary, causing a
temperature jump of $\Delta T\sim 300$ K across it,
nearly restoring pressure equilibrium, albeit at the expense of  
throwing the outside gas out of thermal equilibrium.
Figure~\ref{den75c}a shows the density, the bulk and sound speeds,
and the cooling times out to $75$ pc from the center 
at $27.0$ Myr. Within the region $1 < |x| <75$ pc from the center, the
density lies in the range $0.5<\rho<3$
cm$^{-3}$, with cooling times between $0.8$ and
$2.7$ Myr. Within $20<|x|<75$ pc, the cooling time is roughly constant
($\tau_{\rm cool}\approx 2.0$--$2.7$ Myr), 
and the typical value of $\eta$ with respect to the size of this region
(55 pc) is $\sim 0.8$--$0.9$. This value of $\eta$ explains the
approximately isobaric evolution of the flow across this region, which
is characterized by relatively low densities ($\rho \lesssim 0.7$ cm$^{-3}$). 
At intermediate
densities, i.e. within $3<|x|<20$ pc, $\eta$ also has values $\sim 1$
(again with respect of the region size, $\sim 17$ pc),
because both $\tau_{\rm cool}$ and $\tau_{\rm s}$ decrease towards
the cloud. Thus, {\it this simulation has evolved towards values of $\eta$
closer to unity over most of the medium external to the cloud}, except
at the cloud core ($|x|<0.25$ pc), where the condition 
$\eta \ll 1$ is preserved (\S
\ref{sec:entr_pert}). Note also that for perturbations significantly
smaller than the sizes of the regions mentioned above, $\eta$ is
actually $\gg 1$ in the warm medium.

The stage of slow accretion at the cloud boundary
is established after $\sim 25$ Myr. In Figure~\ref{den75c}b the
central part of the simulation, including the accretion front, is shown
at $t=27.0$ Myr. It is worthwhile to note that there is still an appreciable
excess of density with respect to $\rho_{\rm isob}$ at that
time ($\rho > 2 \rho_{\rm isob}$). The column density of
the cloud is $1.65$ M$_{\odot}$ pc$^{-2}$ at $27.0$ Myr, 
very similar to the initial column density within
$|x|<\sigma$ ($1.68$ M$_{\odot}$ pc$^{-2}$), whereas 
the mass column density remaining in unstable regions
with $\rho< 3$ cm$^{-3}$ and $1400<T<6000$ K (not including the region
unaffected by the perturbation) is $3.2$ M$_{\odot}$ pc$^{-2}$. 
In fact, only a small
fraction of the low-density gas has reached densities corresponding to
the warm stable phase (Fig.\ \ref{den75a}) (no gas at all reached that
phase in runs DEN3 and DEN15). As mentioned with respect to run DEN3,
this is in part a consequence of having considered a single density
perturbation in an unstable, homogeneous medium. 

The final slow-growth state of this large-scale perturbation is actually
very similar to that of DEN3, after the shock has established a
quasi-isobaric regime. The gas still accreting has subsonic velocities,
is out of thermal equilibrium, and is traversing the ``unstable''
density range, although with density and pressure fluctuations of the
same sign, so that it has no further tendency to fragment. Due to the
low accretion rate, long times will be required to evacuate the unstable 
range, as already pointed out by Goldsmith (1970).

\subsubsection{Convergence study} \label{sec:convergence}

\begin{figure}
\plotone{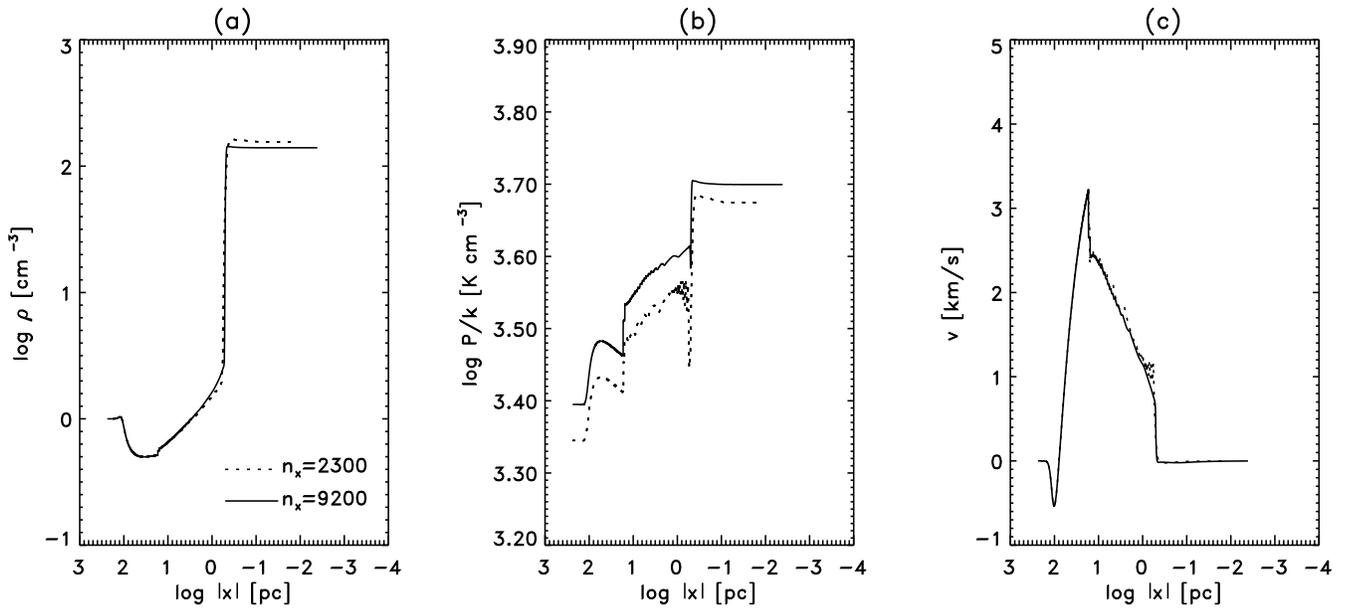}
\caption{
Comparison of density (panel a), pressure (panel b) and velocity 
(panel c) for run DEN75 with
different resolutions. In panel (b) the pressure for the low-resolution
simulations has been shifted $0.05$ to make the plot readable.
}
\label{res_stud}
\end{figure}

In order to assess the effect of numerical resolution, we have
conducted three realizations of each model DEN3, DEN15 and DEN75b, in
which the resolutions differed by factors of two\footnote{We present
here the case of DEN75b because it is the most computationally
demanding situation.}.
For DEN3, the number of grid points, $n_{x}$, were $800$, $1600$ and
$3200$, which correspond to a resolution around $x=0$
of $3\times 10^{-2}$, $1.5\times 10^{-2}$ and $7.6\times 10^{-3}$
pc, respectively. Almost identical results were found (differences of less
than $1\%$ in all variables).
For DEN15 we used $1800$, $3600$ and $7200$ grid points, with associated
resolutions $7.6\times 10^{-2}$, $3.8\times 10^{-2}$ and $1.9\times
10^{-2}$ pc, respectively. Again, the variations were insignificant
(less than $1\%$). 
For DEN75b, simulations were conducted with $2300$, $4600$ and $9200$
grid points. In Fig.~\ref{res_stud}
we compare two snapshots for $n_{x}=2300$ and $n_{x}=9200$ at an advanced
time when steep gradients are already present. The density and
velocity profiles are similar in both cases, whereas the
pressure is slightly less smooth in the low resolution simulation.
In the high-resolution simulation the pressure is better
resolved but the velocity field is unaffected. 

It is possible that the growth of small-size structures consisting of a few
grid points are limited by resolution. To be sure that the results
are not sensitive to resolution, experiments with increasing resolution
have been carried out for all of the simulations
presented throughout this paper. High enough resolution has been used
to ensure that it does not affect the results. See also Paper III for 
further discussion.

\subsection{Multi-scale density perturbations} \label{sec:scale_compet}

As mentioned above, interesting phenomena may be expected if the
initial density
profile contains a spectrum of wavelengths (see also Burkert \& Lin
2000). Small scale perturbations ($\eta > 1$) have larger growth rates
than larger-scale ones ($\eta < 1$) and thus will quickly outgrow them.
Suppose now that we have two separate scales, one with
$\eta\ll 1$ and the other one $\eta\gg 1$. If the clouds
formed from small scales do not consume an important
fraction of the mass, their bulk motions will
be dominated by the pressure structure of the larger scales.
As a consequence, these clouds will behave as if passively advected by
the larger ones and will be accelerated towards pressure minima.
In this case the evolution
may be envisaged as a ``linear'' superposition of the
evolution of small and large scales. In a more general situation,
however, the evolution could become extremely complex because of
the competition between the growth of perturbations of different sizes.
To  quantify the relevance of this competition on the dynamics and
mass spectrum of the forming clouds, we consider
a superposition of fluctuations
with wavenumbers in the range $k_{\rm min}<k<k_{\rm max}$,
where $k_{\rm min}=2\pi/l$, with $l$ the box size, and
$k_{\rm max}$ is varied to consider various extents of the perturbed
range of scales. The amplitudes of the fluctuations
follow a power-law of the form $\rho_{k}=C_{0}(k/k_{\rm min})^{-\delta}$.
We consider two values of the ratio $k_{\rm max}/k_{\rm min}$ and two
values of the spectral slope $\delta$. Moreover, we consider another set
of simulations in a box of size 10 times smaller. Note that, if the
amplitude of the initial perturbations is an
increasing function of the wavenumber (i.e. $\delta <0$),
it is clear that the small scales will be able to consume an important
fraction of the gas with no chance for large scale perturbations
to survive.  Thus, we restrict our discussion to $\delta >0$, in
particular to the values 3/2 and 1/2. 
For these values of $\delta$ we now discuss the conditions under which
the small-scale perturbations do not appreciably affect
the overall evolution.

\begin{figure}
\plotone{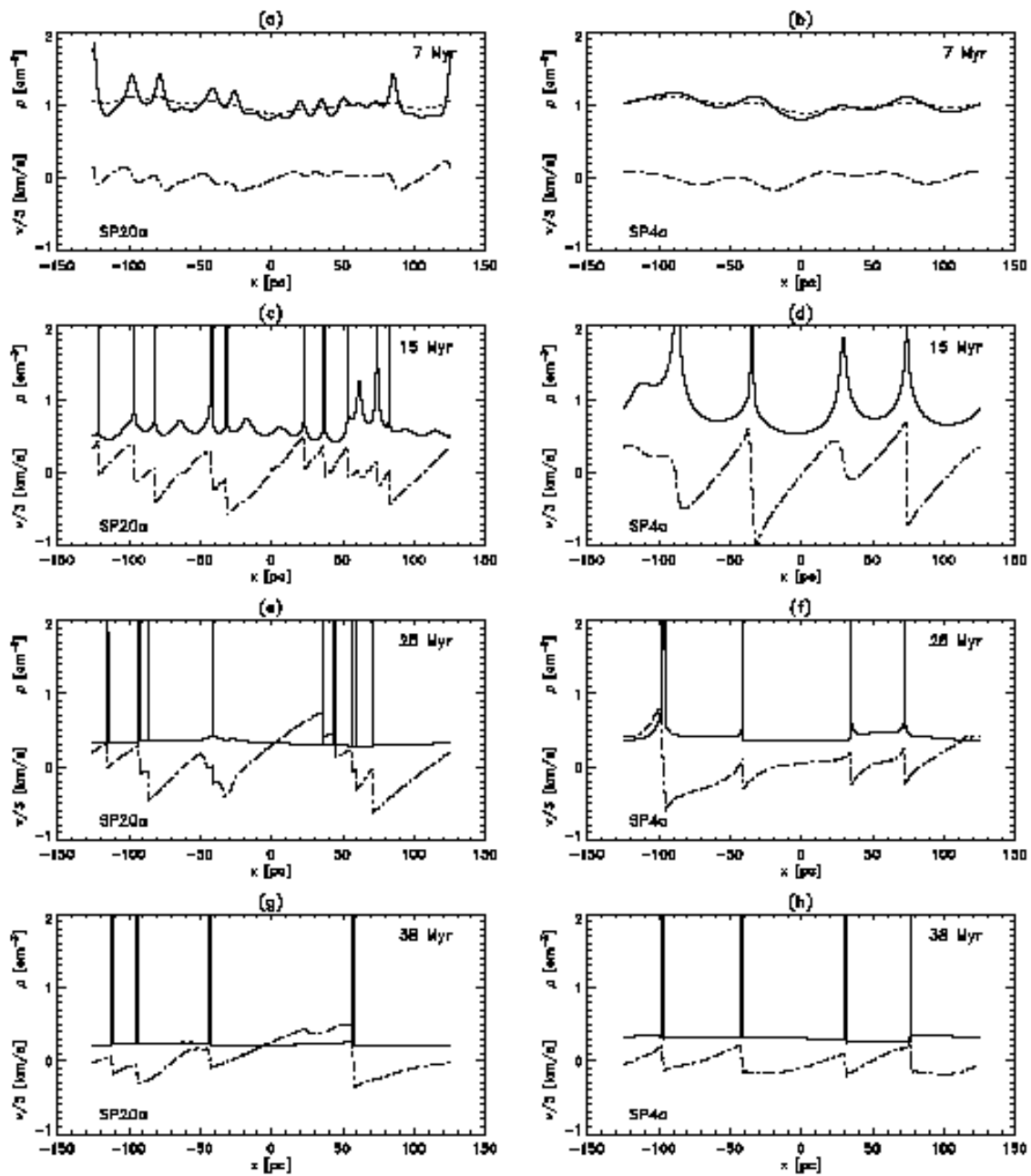}
\caption{
Density ({\it solid line}) and velocity (divided by 3) 
({\it dot-dashed line}) 
profiles at different times for runs SP20a ({\it left panels}) and SP4a
({\it right panels}). 
In the upper panels the initial density configuration is also plotted 
({\it  dotted line}). Clouds undergo an overshoot of density in the
crushing stage (not shown). Most of the fully-condensed clouds present
densities between $40$ and $80$ cm$^{-3}$ but can be larger when two clouds
collide.}
\label{sp20l}
\end{figure}

First, we compare the evolution of two runs with $\delta=3/2$, labeled
SP20a and SP4a (for
``superposition''; see Table~\ref{T1}). These runs
have ratios $k_{\rm max}/k_{\rm min}=20$ and 4, respectively. We
consider a box size of $250$ pc, which
implies that, for these runs, $\eta_{0}(k_{\rm min})\simeq 0.04$.
Four snapshots of their evolution are displayed in Figure~\ref{sp20l}.
The number of condensations at intermediate times
is $16$ and $5$ for SP20a and SP4a, respectively.
This number depends almost linearly on $k_{\rm max}/k_{\rm min}$
for these simulations.

The problem of competition for the available mass can be summarized in
the question of whether clouds formed from small scales in run SP20a group
together to form a density pattern that at large scales is similar
to that of run SP4a. If that happens, we can say that perturbations at small
scales are unable to suppress the development of TI at larger scales.
Note that we speak in terms of Fourier wavelengths only for the initial
spectrum. In contrast, once condensations are growing, we focus
on their clustering and distribution in physical space.

 From panels (c) and (e) of Figure~\ref{sp20l},
we see that some of the incipient
condensations in run SP20a are disrupted when they are
embedded in a diverging flow (for instance the
clumps between $x=-25$ and $x=20$). Conversely, structures
placed in a compression zone grow very fast.
Note that some of the
clouds in the group near $x=60$ are moving relatively fast
($\sim 1.5$ km/s) at $25$ Myr.
At later times, the clouds undergo a series of
mergers that ultimately lead to four main condensations at positions
$x=-110$, $x=-95$, $-40$ and $60$ at $t \approx 38$ Myr.

For this run, the column density ratios of gas mass
between the warm,
the thermally unstable and the cold phases are $1:0.5:1$ at
$15$ Myr and quickly evolve to $0.69:0.025:1$ at $20$ Myr, whereas
at $25$ Myr they are $0.46:0.01:1$.

As expected from their larger sizes, the growth rate of the density structures
is smaller in run SP4a, but each condensation accretes more matter.
In fact, most of the gas is still in the thermally unstable phase
at $15$ Myr, with column density ratios $0.9:9.7:1$.
At times $\sim 25$ Myr, these ratios are $0.66:0.034:1$. Still, the
fraction of gas in the thermally unstable phase is higher than
in SP20 at $20$ Myr.
At those times ($\sim 25$ Myr), four main condensations have already
formed at $x=-100$, $-40$, $35$ and $70$.
Comparing with SP4a, the evolution of run SP20a is somewhat more
dynamical, in the sense that clouds acquire larger
velocities and are prone to suffer mergers more
easily than in run SP4a, in which clouds are more massive and static.
The locations of the condensations are the same except for
those at $x=35$ and $x=70$ which have not had
time to merge yet; they will do so in an exceedingly long
time, at $t\approx 90$ Myr. We conclude that the final states are
not equivalent in this case.

\begin{figure}
\plotone{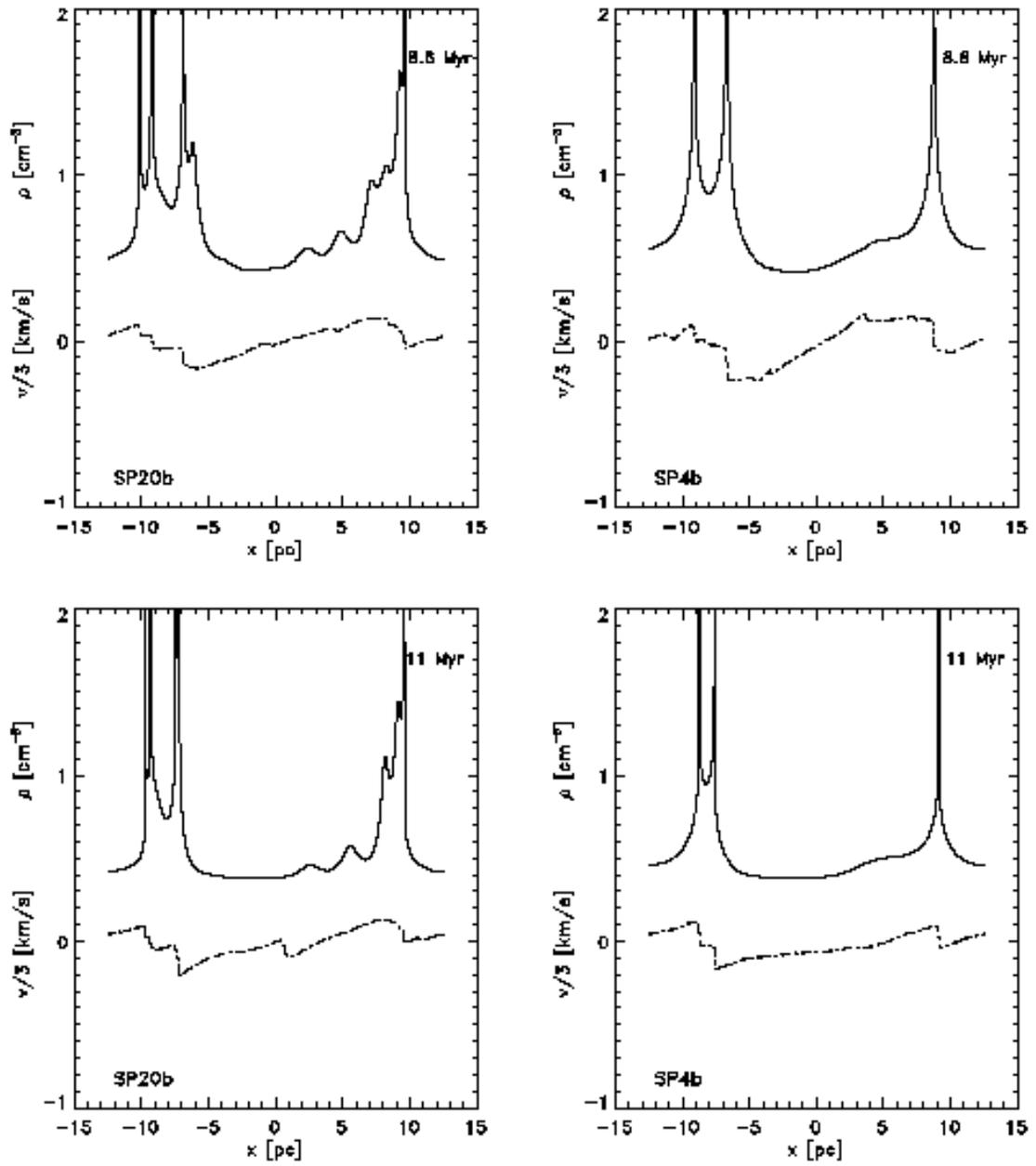}
\caption{
Same as Figure \ref{sp20l} but for runs SP20b ({\it left panels}) 
and SP4b ({\it right panels}). 
}
\label{sp20s}
\end{figure}

Next, we consider two more sets of simulations using the same
density spectrum, but varying the simulation box size.
When the box is five times larger, the differences are even more
significant. However, if the box size is reduced a factor ten
($25$ pc, runs SP20b and SP4b, respectively), so that for
$\eta(k_{\rm min})=0.4$, the final states of the two initial spectra
are much more similar, as seen especially from their velocity fields 
(see Fig.~\ref{sp20s}). Hence, the evolution is not only sensitive to the 
initial spectrum of the perturbations, but also to $\eta(k_{\rm min})$.

\begin{figure}
\plotone{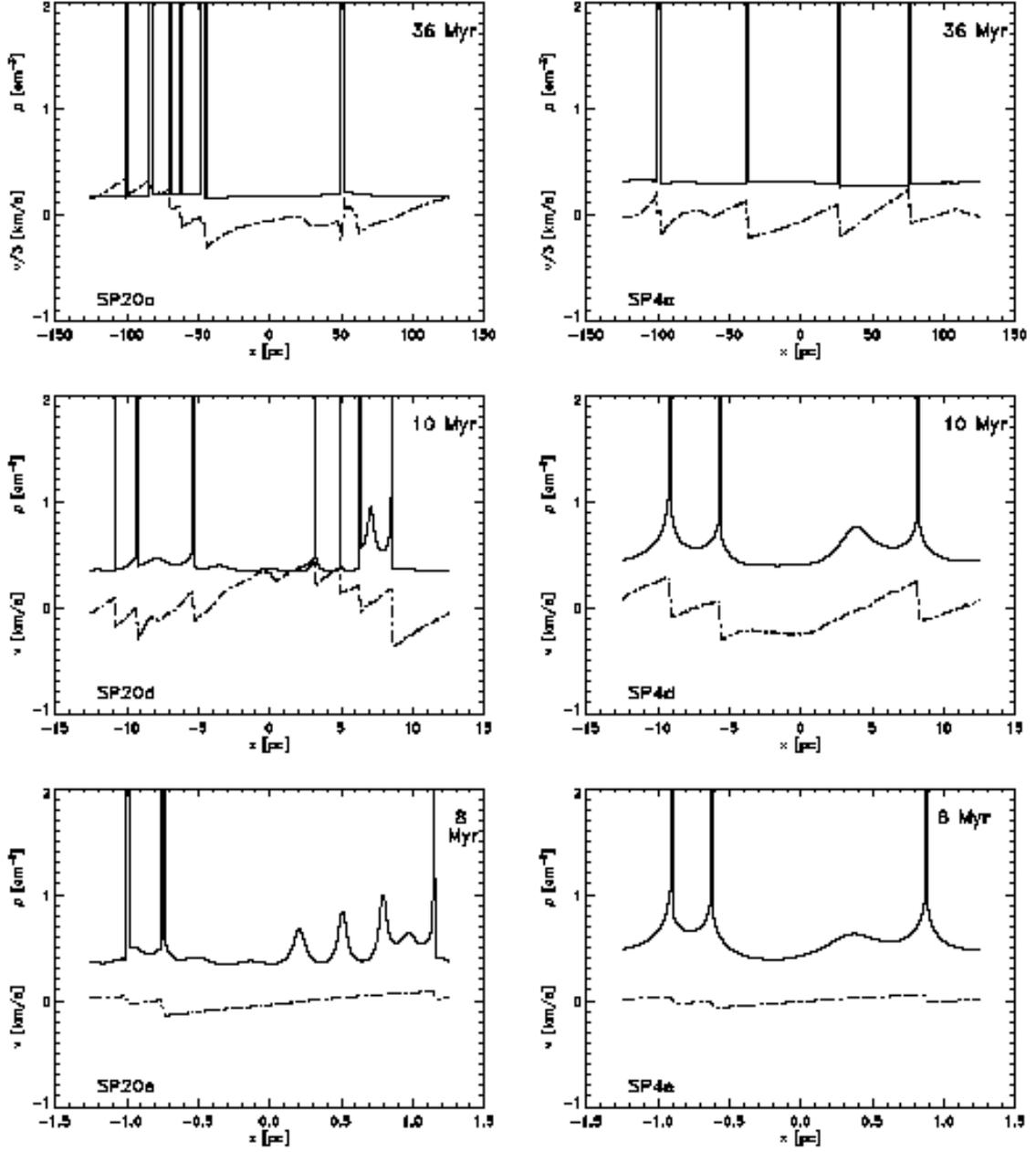}
\caption{
Density ({\it solid line}) and velocity  
({\it dot-dashed line}) 
profiles for runs SP20c,d,e ({\it left panels}) 
and SP4c,d,e ({\it right panels}). 
Note that for spectra with $k_{\rm max}/k_{\rm min}=4$ the evolution
proceeds more slowly.}
\label{sp20c}
\end{figure}

Let us now consider a set of similar experiments
but with $\delta=1/2$. First, we take $\eta_{0}(k_{\rm min})=0.04$
(runs SP20c and SP4c). In Figure~\ref{sp20c} we see that
run SP20c contains several condensations that
cannot be associated to any main cloud in SP4c.
The most noticeable difference between SP20c and SP4c is that
SP20c is able to group condensations into two clusters whereas
the four condensations in SP4c would need an exceedingly long
time to coalesce, as can be seen by comparing the velocity
gradients between positions $x=-110$ pc and $x=-50$ pc
for the two runs (Figure~\ref{sp20c}). Hence, the final states of runs
SP20c and SP4c are qualitatively very different.

Next, we consider
a set of runs with $\delta=1/2$ and $\eta_{0}(k_{\rm min})=0.4$ (runs
SP20d and SP4d) and another set with $\eta_{0}(k_{\rm min})=4.0$ (runs
as SP20e and SP4e). Run SP20d still exhibits noticeable differences with
respect to SP4d. In Figure~\ref{sp20c} we see that there are
three condensations between $x=1$ and $x=7$ pc, generated
from small scale perturbations in SP20d, but they are absent in
SP4d. These condensations have positive velocities,
whilst the gas between $x=1$ and $x=5$ has
the opposite velocity in SP4d.
This implies that in this case the emergence of small-scale
condensations produces important
modifications on the evolution of the gas at large scales because
the late-time mass distribution is significantly altered if
they are present. However, the opposite conclusion
is reached for SP20e and SP4e, which are seen to have more similar
states, again judging in particular from their velocity fields (see
Fig.~\ref{sp20c}). The density peaks between $x=0$ and $x=0.75$ pc
in run SP20e do not grow at all between $6$ and $8$ Myr.

We conclude that both the spectrum exponent and $\eta (k_{\rm min})$ 
determine the outcome of a superposition of modes of various scales. It
appears that, as the initial density power spectrum is shallower, the
perturbations need to extend to smaller scales (larger $\eta$'s) in
order for the small-scale fluctuations not to strongly disturb the
growth of the larger-scale ones. In particular, we have found that, for
$\delta = 3/2$, a value $\eta (k_{\rm min})=0.4$ is required, while for
$\delta = 1/2$, a more stringent condition, $\eta(k_{\rm min})=4.0$ is
necessary. 
These critical values could be different for 2D or 3D systems.

It is worthwhile to point out that even though the development of TI may
be suppressed at large scales due to mass consumption by small
scales, massive clouds may form by coalescence of small clouds
(see e.g., Elmegreen 1990, Murray \& Lin 1996).
As mentioned before, clouds formed at small scales may present
appreciable bulk velocities
that facilitate the merger of clouds. This effect has recently been seen 
also in 2- and 3D simulations by Koyama \& Inutsuka (2002) and Kritsuk
\& Norman (2002). The coalescence is exaggerated in
one dimensional systems, because the merger
rate does not depend on the filling factor of the clouds. However, in
more than 1D, the merger rate may still be large if the small-scale clouds are
advected towards the condensation center of the large-scale ones (see \S
\ref{sec:lim_1d}). In general,
our 1D simulations show that small scale perturbations
do not lead to a ``forest'' of small clouds because they
coalesce, building more massive ones.

\subsection{Simulations of forced flows} \label{sec:vel_compet}

We now consider the case of flows forced continuously in time,
representative of a turbulent ISM, to determine whether turbulence
may oppose the condensation process (cf., \S \ref{sec:vel_fluc}) under
some circumstances, and, if so, at which scales and values of the rms
Mach number. 
In a domain of size $l$, we force the flow at scales large 
compared to the final sizes of the condensations. 
This implies that, once formed, the condensations cannot be disrupted by
the random forcing. Note, however, that in higher
dimensionalities dynamical instabilities could disrupt the clouds while
they form, or after they have formed; see \S \ref{sec:lim_1d} for  
further discussion. 

We adopt a random force ${\mbox{\boldmath $f$}}$
of the form 
\begin{equation} \label{eq:forcing}
{\mbox{\boldmath $f$}}(x,t)=\hat{{\mbox{\boldmath $x$}}}
\Re\left\{N \exp \left[ik\left(t\right) x+i\phi\left(t\right)\right]\right\},
\end{equation}
where $\Re$ denotes the real part, $k(t)$ is a time dependent
wavenumber, $\phi(t)$ is the 
phase and $x$ the position. Following Brandenburg (2001), we
take $N=f_{0}c_{\rm s}\left[k(t) c_{\rm s}/\delta t\right]^{1/2}$,
where $f_{0}$ is a constant factor and $\delta t$ is the length
of the timestep. The values of
$k/(2\pi/l)$, where $l$ is the box length, and of $\phi$ are selected at
each timestep randomly in the ranges $\left[3,10\right]$ and
$\left[0,2\pi\right]$, respectively. This implies that the scale of
maximum forcing is $\lambda_{\rm f}=l/10$.
The positive exponent (1/2) in $N$ implies 
that the maximum strength of the forcing occurs at the highest
wavenumbers $k$, i.e., roughly at 1/10 of the box size. A summary of the
forced simulations is given in Table~\ref{tab:T2}.
There, $\tau_{\rm cond}$ is defined as the time for
a condensation to reach a density $\rho=10\rho_{0}$ for the first time,
$\left<e\right>$ is the ratio of the time-averaged internal energy (with the
average performed over $\tau_{\rm cond}$) to its initial value.
${\mathcal{M}}_{\rm rms}$ is the rms Mach number, also averaged over
$\tau_{\rm cond}$ to increase statistical confidence. Note
that ${\mathcal{M}}_{\rm rms}$ 
fluctuates strongly in time, in contrast to $e$, which behaves much
more smoothly. The different values of 
$\tau_{\rm cond}$ in one row in Table~\ref{tab:T2} correspond to runs with
different seeds for the random numbers.

Let us first consider experiments with $l=3$ pc, in which $\lambda_{\rm
f}=l_{\rm eq}/30$, since $l_{\rm eq} \sim
10$ pc (cf.\ \S \ref{sec:entr_pert}). For low Mach numbers 
(say ${\mathcal{M}}\leq 0.1$), it then holds that
the sound speed is larger than the typical values of the velocity
fluctuations, with $\eta \gg 1$. Consequently, the condensations behave
essentially as entropy perturbations, and evolve
approximately isobarically (cf.\ \S \ref{sec:quali}).
Interestingly, however, the various realizations of run 1, which have
$l=3$ pc and ${\mathcal{M}}=0.03$,
present a mean $\tau_{\rm cond}\approx 12$ Myr, about three times
the condensation time for the case with an initial density perturbation
of amplitude $20$ \% and the same size. A large dispersion
of $\tau_{\rm cond}$ is apparent for flows with ${\mathcal{M}}_{\rm rms}
\approx 0.15$--$0.3$, suggesting that the formation of condensations
is sensitive to the past history of the flow, becoming somewhat fortuitous
for higher Mach numbers.
For even larger values of ${\mathcal{M}}_{\rm rms}$ ($\gtrsim 0.3$),
eventually the condensation is suppressed altogether (runs 5--9). In
these cases, the internal energy initially increases with time until it
starts to fluctuate ($\sim 4\%$) around a value more than twice its
initial value. 
This occurs when cooling is able to compensate
the thermal energy injection by the random forces, since the net
cooling rate increases also with time as both the strength
of the perturbations in density and temperature increase. 
Specifically, energy injection and net cooling are in balance with
a mean value of $0.038$ erg s$^{-1}$ g$^{-1}$ in run 7 
after a time $\sim 0.8$ Myr,
whereas this occurs after a time $\sim 0.35$ Myr and for a value of 
$0.175$ erg s$^{-1}$ g$^{-1}$
for run 9. 
Our results thus suggest empirically
that, for the cooling functions we use, the transition to stabilization occurs
roughly at $e\approx 2$ (${\mathcal{M}}_{\rm rms} \sim 0.3$). 

On the other hand, for the case $l=100$ pc (run 17), it is clear that the
forcing is applied at scales where $\eta \lesssim 1$, and indeed the
density fluctuations generated by the random compressions rapidly
cool and acquire a
temperature and a pressure close the thermal equilibrium values. Since the 
average density is in the unstable range, TI proceeds unimpeded,
manifesting itself not so much as a tendency for the fluctuations to
grow, but in the null resistance of the flow to external compression (\S 
\ref{sec:vel_fluc}). Indeed,
the random compressions {\it promote} the instability, as indicated
by the decrease in $\tau_{\rm cond}$ with increasing ${\mathcal{M}}_{\rm
rms}$. This is the regime of the randomly forced simulations of Paper I, and
corresponds to the case of an effective imaginary sound speed of the
wave mode, which consequently does not propagate (e.g., Field 1965; Shu
1992). 

The case with $l=10$ pc (runs 10--16) 
has $\lambda_{\rm f}\sim l_{\rm eq}/10$. Nevertheless, Table
\ref{tab:T2} shows that the condensation of the density
fluctuations cannot be inhibited anyway, even 
for ${\mathcal{M}}_{\rm rms} \sim 0.6$. We see that $e$ can hardly 
be increased by factors of $1.9$ in this case. We understand this
as a consequence of the fact that, for such large Mach numbers,
the reduction of the cooling time by the compression-driven density
enhancement can reduce $\eta$, allowing the triggering of the
instability under the isobaric mode by the nonlinear velocity
fluctuations. It appears that this does not happen in
the case $l=3$ at large ${\mathcal{M}}_{\rm rms}$ because $\eta$ is
large enough in that case that the shocks produced are too close to adiabatic
and cannot raise the density enough to destroy the adiabaticity.  
We should remark that for simulations with $l=10$ pc and strong forcing 
(${\mathcal{M}}_{\rm rms} > 0.5$, e.g.~run 16)
compressions can produce transient density fluctuations $\sim 10\rho_{0}$, 
which are dispersed by their own pressure.

We conclude that, as conjectured in \S \ref{sec:vel_fluc}, the growth of 
perturbations {\it can} be suppressed provided that the forcing is 
applied at scales small enough that $\eta \gg 1$
and produces a large enough rms Mach number so that $\eta$ is 
dominated by the turbulent velocity fluctuations rather than by the
sound speed. This is because, under these conditions, the perturbations
are nearly adiabatic, and are (at least) linearly stable.
The fact that even large rms Mach numbers in runs with $l=3$
pc do not induce condensations indicates that the stability of these
perturbations goes well beyond the first order.

\section{Discussion} \label{sec:discussion}

\subsection{Comparison with previous work} \label{sec:comparison}

Some of the issues considered in this paper have been discussed by
previous authors in the context of the fragmentation of proto-galactic
and proto-globular cluster clouds (e.g., Murray \& Lin 1989, 1990),
cooling flows in galaxy clusters (e.g., David et
al.\ 1988; Brinkmann et al.\ 1990; Malagoli et al.\ 1990; Kang et al.\
1990) and the ISM (e.g., Hennebelle \& P\'erault 1999, 2000; Burkert \&
Lin 2000; Koyama \& Inutsuka 2000). In this section we
put our results in perspective with respect to those works. In Paper I we have
presented a comparison between our ISM-oriented scenario and previous
work considering quasi-static development of TI.

It should be first pointed out that some of those studies have
solved the chemical rate equations, allowing for
out-of-chemical-equilibrium abundances of the
relevant coolants (e.g., Murray \& Lin 1989,
1990; David et al.\ 1988; Kang et al.\ 1990; Koyama \& Inutsuka 2000).
This is crucial in situations in which considering
out-of-chemical-equilibrium cooling rates can make the difference
between stability and instability. However, it is not so crucial
in relation to the present-day ISM, which is already generally agreed to 
be thermally unstable in the range 300 K $\lesssim T \lesssim$ 6000 K
under chemical equilibrium conditions. Once the medium is
known to be unstable, it seems reasonable
to think that variations in the cooling rates due to deviations
from chemical equilibrium should only cause modifications in the details
of the condensation process, but that the main qualitative 
behavior should not change significantly. For example, the
development of supersonic motions by the development of TI similar to
our results in \S \ref{sec:ev_density} in the
context of cooling flows and of proto-globular cluster clouds has been
observed by David et al.\ (1988) and Kang et al.\ (2000), respectively.
Moreover, Murray \& Lin (1989, 1990) discussed the effects of various
types of perturbations, cases with and without background heating, and
the need for a minimum cloud collision velocity (equivalent to a
compression) in order to trigger TI in proto-globular cluster clouds
subject to external heating. The latter result is similar to that of the 
more recent study by Hennebelle \& P\'erault (1999) in the context of
the ISM, even though these authors did not explicitly solve for the
chemical rates. Thus, similar phenomenology is found with
or without non-equilibrium chemistry, once the system is unstable.
In any case, the caveat of our neglect of
non-equilibrium chemistry and rates should be kept in mind.

Now let us specifically consider recent studies of TI in the context of
the ISM. First, as mentioned above, the 
studies of Hennebelle \& P\'erault (1999) and Koyama \& Inutsuka (2000)
were again concerned with the triggering of TI in the {\it stable} phases of
the ISM by means of strong shocks or compressions. These studies
are related to the issues we discuss in \S\S\ref{sec:vel_fluc} and
\ref{sec:vel_compet}, although in our case the medium is already
isobarically unstable from the start. Interestingly, however, the case
of velocity perturbations with $\eta \gtrsim 1$ is somewhat resemblant
of the case of an initially stable medium, in the sense that a strong
enough compression is needed in order to raise the cooling rate to high
enough values that condensation can occur. Also,
it is possible that the induction of TI in the cold neutral medium due
to molecular cooling found by Koyama \& Inutsuka could occur during the
transient overshoot we found in the condensation of large-scale
perturbations if we included such cooling, but in any case this
phenomenon would only support our suggestion that the static
pressure-equilibrium configuration is hard to achieve, since the diffuse 
clouds could be subject to a further instability that would transform
them all the way to small molecular cloudlets.

On the other hand, Burkert \& Lin (2000) did aim their study at the
formation of clumpy clouds in an unstable ISM, but they considered a
different physical situation than the one we have
considered here, a fact which caused them to reach significantly different
conclusions from us. Specifically, they considered a case with no background 
heating, a choice that changes the instability criterion and, more
importantly, implies that the whole flow is cooling, so that the
development of TI consists of a perturbation cooling faster than the 
background medium. Thus, perturbations that cool isochorically and grow
slowly (i.e.,
large-scale ones) cannot get ``too far ahead'' of the background gas
before a minimum temperature is reached at which the gas either ceases
to cool or exits the unstable range. As a consequence, Burkert \& Lin
concluded, similarly to Murray \& Lin (1990) in the context of
protogalactic clouds, that 
such perturbations cannot reach very large density enhancements
over the background, and that only isobarically cooling ones do. This is
in sharp contrast to our conclusion that it is precisely the large-scale 
fluctuations that reach the highest amplitudes (albeit only
transiently). The reason is that in our case the background is not
cooling, and thus there is no competition between it and the
fluctuations to reach the minimum temperature. Which case applies at a
given location in the real ISM should depend on the presence or absence
of a population of heating sources that maintain a relatively constant
background heating rate. It is in general agreed that the ISM is
permeated by a roughly uniform UV field and cosmic rays that provide a
background heating, although non-negligible variations are certainly
expected to occur (Hollenbach, Parravano \& McKee 2001).

\subsection{Adequacy of the one-dimensional approach and two-dimensional 
simulations}
\label{sec:lim_1d} 

The one-dimensional numerical approach that we have used here clearly
has some limitations. It was chosen in order to fully resolve the
condensations even at the time of maximum compression, but it has the
disadvantage of being unable to capture effects that involve solenoidal
(vortical) motions in the flow. In particular, it has been suggested
that dynamical instabilities such as Rayleigh-Taylor (RT),
Ritchmyer-Meshkov (RM), or Kelvin-Helmholtz (KH) may
destroy the clouds during or after their formation in stratified media
(e.g., Malagoli et al.\ 1990; Hattori \& Habe 1990; Reale et al.\
1991; Murray et al.\ 1993; but see Vietri et al.\ 1997 for an opposite
view). However, within the scope of the present paper, in which we have
chosen to study the development of TI in isolation, we are not concerned
about the future fate of the clouds formed, but only about issues of the
cloud formation process within the idealized scenario that TI is
responsible for this process. This amounts to studying the limiting,
most favorable case in which the classic scenario of an equilibrium ISM,
with thermal-pressure bounded clouds and sharp phase transition could
arise, and our main conclusion is that even within this scenario such a
picture of the ISM appears unlikely. 

Perhaps the closest point of contact of the issues discussed in this
paper with the possibility of cloud destruction by dynamical
instabilities occurs in the context of our discussion of the evolution
of velocity perturbations. As we have seen, these may avoid
condensation for arbitrarily long times if they have moderately large
amplitudes and occur at small enough scales. Thus, in a sense, the velocity
fluctuations even in our simple 1D case have the effect of counteracting the
development of the isobaric mode of TI analogous to that of the dynamical
instabilities discussed by other authors.

It is thus of interest to compare the effects of those
dynamical instabilities on the condensed clouds with the effects of the
random forcing we have considered here. Concerning the KH instability,
it is expected to occur as the condensations begin to sink
in a stratified medium as they develop inverse buoyancy (Balbus \&
Soker 1989). Thus, it could only start to act after the condensation has 
reached a sufficiently advanced evolutionary stage. That is, its effect
is one of destroying a cloud after it has formed, rather than
obstructing its growth, as we have found here.

On the other hand, the RT instability might indeed occur while the
condensation proceeds, as the flow is accelerated inwards of the
condensation, with the density gradient pointing also inwards. To test
this possibility, we have performed two two-dimensional simulations
analogous to run DEN75, but including weak random initial velocity
perturbations with rms Mach numbers 0.1 and 0.01 with a half-wavelength
of 1/8 of the box. We find no traces of development of RT. Instead, the
velocity fluctuations generate density fluctuations that evolve in a
very similar way to the simulations of \S
\ref{sec:scale_compet}, with the small-scale fluctuations condensing
first, and then merging towards the center of the large-scale
perturbation. The certainty that RT does not arise comes from the fact
that the velocity field converges onto the fragments rather than shear
between them, as would be expected for RT. Thus, we conjecture that TI
development inhibits that of RT. 

Finally, the RM instability probably does not apply to single clouds
formed by TI because it occurs upon the interaction of a shock with a
discontinuity. Instead, what we have found in this paper is that the
evolution of a single perturbation {\it generates} a shock wave that
propagates outwards from the condensation. It may be relevant, however,
in the case of multiple fluctuations, as in \S \ref{sec:scale_compet},
in which the shock propagating out of one condensation may interact with 
another. However, this again refers to the future evolution of the
condensations formed by TI, which is out of the scope of this paper.

A final note is that, in any case, the effects of all these
instabilities go in the same direction as the issues discussed in this 
paper: as they tend to destroy the condensations, they tend to restore a 
more continuum-like ISM, rather than one separated in discrete phases.

\subsection{Implications} \label{sec:implications}

The results of this paper suggest a number of implications. All of
them require further investigation, but the possibility of their
realization in the ISM is of great interest:

1. The outcome of the development of TI, even if it
were the sole cloud-forming agent in the atomic ISM, is not
necessarily the production 
of quiescent thermal pressure-bounded clouds in contact with the warm
stable phase. Large-scale density
perturbations require long times to condense (over 7 Myr for
perturbations of size $> 10$ pc), during which they are in a highly
dynamic state, far from pressure equilibrium. At later stages,
pressure equilibrium is approximately restored, but at the expense of
disrupting thermal equilibrium of the surrounding medium through the
shock propagating away from the condensation. This surrounding gas
continues to accrete onto the condensation for much longer times,
providing a ram presure that keeps the cloud density above the
pressure-equilibrium one by factors of up to 2. Moreover, the accreting
gas is not in thermal equilibrium, but in a nearly isobaric regime, 
in which the (small) pressure fluctuations have the same sign as those of the
density, and therefore the medium has no strong tendency to fragment any
further. In particular, this non-equilibrium situation implies that the
unstable density range as shown in Fig.\ \ref{cooling} does not exactly
coincide with the unstable temperature range.

Smaller-scale (large-$\eta$) fluctuations ($\lesssim 3$
pc at $\langle\rho\rangle \sim 1$ cm$^{-3}$, or even larger at
proportionally lower densities) can condense and reach near
thermal-pressure balance in shorter times ($\lesssim 4$ Myr) if they are 
nearly isobaric in nature, but they are stable if they are nearly
adiabatic, a condition that can be accomplished if they originate from
velocity fluctuations with sufficiently large ${\mathcal M}_{\rm rms}$
at sufficiently small scales (\S \ref{sec:vel_compet}). Moreover,
small-scale fluctuations, if part of a full spectrum of 
fluctuations, may continue to form part of the condensation of
large-scale ones, if the relative amplitude of the former is small
enough and they occur at sufficiently large $\eta$
(\S \ref{sec:scale_compet}).
In general, it appears that the pure nonlinear development of TI, 
even in the absence of other agents like magnetic fields, rotation,
stellar energy injection, etc., does not necessarily lead to rapid
formation of clouds with sharp phase transition at their boundaries.

2. The existence of accreting gas in the ``unstable'' range at late
stages of the evolution, and the stabilization produced by velocity
fluctuations in a certain range of scales suggests that in the
turbulent ISM the ``unstable'' phase may actually be significantly
populated, since the density fluctuations are in general expected to
have a dynamical origin. This may explain observational (e.g., Dickey,
Salpeter \& Terzian 1977; Kalberla et al.\ 1985, Spitzer \&
Fitzpatrick 1995, Fitzpatrick \& Spitzer 1997; Heiles 
2001) and recent numerical results by Gazol et al.\ (2001) and
Kritsuk \& Norman (2002),
suggesting that relatively large amounts of gas are present in the ISM
in the unstable temperature range. 

3. In summary, both the gas accreting onto the condensations at late times
and the ``unstable'' gas in the presence of {\it stable} adiabatic
fluctuations, are out of thermal equilibrium. In general, it can be said 
that {\it thermal and pressure equilibria are mutually exclusive for gas
in the unstable range}, because each implies a different 
value of the pressure at densities intermediate between $\rho_0$ and
$\rho_{\rm isob}$ (fig.\ \ref{fig:isob_vs_eq}). This incompatibility between
the two equilibria can only be avoided by means of sharp phase
segregation, i.e., when no gas in the unstable exists. However, if such
gas exists, it should in general be expected to be out of thermal
equilibrium. This suggests a possible observational test for determining
whether the gas apparently seen at unstable temperatures corresponds to
this kind of regime, or at least whether it is in thermal equilibrum or
not.  This could be done by either a) simultaneously determing two of
its thermodynamic variables, or b) comparing directly observed 
cooling rates (e.g., fine structure lines) with theoretical estimates 
of the heating rate (e.g., photoelectric heating) in specific
regions (C.\ Heiles, J.\ Scalo, private communications). If this were
confirmed, it would provide strong evidence
that turbulent motions populate all regions of the thermodynamic
variable space, preventing a sharp segregation of the atomic ISM into
the stable phases of TI.

4. The nearly adiabatic response of the gas to velocity perturbations
in cases with relatively large values of $\eta$, suggests that the
warm diffuse ISM, in which this condition is satisfied with respect to
perturbations of up to several pc (\S \ref{sec:ev_density}), may exhibit a
very weakly compressible behavior, and thus be close to a Kolmogorov
regime, which applies to incompressible turbulent flows. This may
explain why observations of the diffuse gas tend to find density
fluctuation spectra with the signature of a Kolmogorov regime (e.g.,
Minter \& Spangler 1996; Stanimirovic \& Lazarian 2001; Dickey et al.\
2001). The numerical results by Kritsuk \& Norman (2002) also suggest a 
nearly adiabatic regime for the diffuse gas.

\section{Summary}\label{sec:summary}

In this paper we have performed a detailed numerical investigation of the
nonlinear development of the isobaric mode of TI in the context of the
atomic ISM, and of its interplay with a spectrum of fluctuations
in density or velocity, aimed at establishing the feasibility of the
classical scenario of a thermal- and pressure-equilibrium ISM, in which
the clouds are directly confined by the more dilute, stable warm gas, as
suggested in the classical two- and three-phase models of the ISM (Field 
et al. 1969; McKee \& Ostriker 1977), 
even under the idealization that TI and a spectrum of density or
velocity fluctuations are the only processes at play.
Most of the discussion is based on one parameter, $\eta$, the ratio
of the cooling to the dynamical times.

We have found that the evolution for initial Gaussian perturbations of
FWHM smaller than $\sim 15$ pc is approximately isobaric. In that case,
a slow-growth stage characterized by quasi-stationary accretion of
low-density gas is
reached in $\approx 4$--$6$ Myr for initial amplitudes of
$10$--$20$ $\%$ over the mean density, although infall velocities $\gtrsim 
0.7$ km s$^{-1}$ still exist at those times. The infall velocities
remain subsonic throughout the evolution for
small perturbation sizes, but infall velocities with Mach number $>0.5$
are acquired if initial perturbations are large enough ($\geq 15$ pc).
For the latter type of perturbations, a shock wave, formed during 
the ``crushing'' and overshooting stage, 
propagates outwards from the cloud and throws the surrounding medium
out of thermal equilibrium, establishing near pressure balance. For the
case with a Gaussian 
density perturbation with FWHM$=75$ pc, the final mass density of the cloud
is more than twice $\rho_{\rm isob}$ due to
the ram pressure of the infalling gas. The time to reach
the slow-accretion stage ranges from 4 to 30 Myr for perturbations with
initial FWHM of 3 to 75 pc. These times are long enough
that a major disturbance is likely to pass through the condensation.

We have quantified the competition between scales by
starting with a power-law spectrum of initial density perturbations,
to assess the possibility that the smaller scales, which have the
fastest growth rates, could consume most of the mass available before
large-scale perturbations have time to grow, thus establishing 
a ``forest'' of thermal-pressure bounded small clouds in relatively
short times. We have found that small-scale perturbations may or may
not use up most of the mass available depending 
on both the spectrum slope $-\delta$ and the initial
values of $\eta$ for the smallest-scale perturbations. 
We have found that the inclusion of large wavenumbers has a progressively
smaller effect on the final state as $\delta$ increases and as the 
perturbations are shifted to smaller scales, increasing $\eta$. For
example, for $\delta = 3/2$, a value of $\eta (k_{\rm min}) > 0.4$ is 
needed in order for the final states to be roughly equivalent, while
for $\delta = 1/2$ this requires $\eta (k_{\rm min}) > 4.0$. 
We looked at this problem to determine whether the dynamical character
of the condensation of large-scale perturbations might be preserved 
even in the presence of small-scale ones. Although our
simulations indicate that the evolution is also very dynamical even
if small-scale perturbations are added, more work is needed to
be conclusive concerning this result.

However, we have found that a much more interesting effect appears in
the case of a medium continuously stirred by an external force.
We have argued that velocity
fluctuations are necessary to produce realistic density fluctuations in
a continuum, their evolution thus ranging from a nearly
adiabatic to a thermal equilibrium response as $\eta$ is varied from
$\gg 1$ to $\ll 1$, provided that in this case $\eta$ is determined by
the turbulent, rather than the sound crossing time. When the
perturbations evolve nearly adiabatically, they cannot condense, because 
they re-expand before they can cool -- i.e., they are {\it stable}. We
have confirmed this suggestion 
numerically, showing that flows with $\langle \rho \rangle = 1$
cm$^{-3}$ forced at scales such that $\eta > 1$ ($\lambda \sim 0.3$--$1$ pc)
and with ${\mathcal{M}}_{\rm rms}\geq 0.3$ do not develop any
condensations over the entire duration of the simulations (between 12
and 26 Myr). However, flows forced at scales such that $\eta \le 1$
($\lambda \gtrsim l_{\rm eq} \approx 10$ pc) produce
condensations regardless of the forcing strength, because in this case
either the effective waves have an imaginary sound speed 
($\eta \ll 1$), and trigger the condensation mode, which is unstable, or
the density enhancement they induce causes
$\eta$ to become smaller than unity by increasing the cooling rate
($\eta \sim 1$).

As implications, we have suggested that:

1. The existence of accreting gas in 
the unstable density range and the stabilization of the flow by
small-scale velocity fluctuations may be at the origin of the relatively
large amounts of gas 
mass in the unstable regime found in both observations (Dickey et al.\
1978; Kalberla et al.\ 1985, Spitzer \& Fitzpatrick 1995,
Fitzpatrick \& Spitzer 1997; Heiles 2001) and simulations (Gazol et al.\
2001; Kritsuk \& Norman 2002) of the ISM.

2. Since transonic velocity fluctuations of scales up to several parsecs
in the warm diffuse ISM have $\eta \gtrsim 1$, this medium should
respond nearly adiabatically to them, and the flow regime should then be 
only weakly compressible. This may explain recent observational results
suggesting that its kinetic energy spectrum 
is very close to the Kolmogorov one (Minter \& Spangler 1996; Dickey et
al.\ 2001; Stanimirovic \& Lazarian 2001), since the latter is expected
to apply to incompressible or weakly compressible cases only.

3. Simultaneous observational determinations of two thermodynamic
variables should allow to distinguish whether the gas apparently seen in the
unstable temperature range is in thermal equilibrium or not, thus
helping to decide whether equilibrium models of the ISM are applicable
or not.

\acknowledgements

We are happy to acknowledge stimulating exchanges with A.~Brandenburg,
C.\ Heiles, P.\ Hennebelle, 
A.~Raga, J.~Scalo and E.~Zweibel. A deep and thorough referee report
prompted much improvement of the paper. This work has received financial
support from CONACYT grants 27752-E to E. V.-S., I32888-E to A. G.,
32139-E to Luc Binette, and a CONACYT postdoctoral grant to F.J.S.S.

\clearpage

\clearpage

\clearpage

\clearpage

\clearpage

\clearpage

\clearpage

\clearpage
\clearpage

\clearpage
\clearpage

\clearpage

\clearpage

\clearpage

\begin{deluxetable}{cccc}
\tablecaption{Summary of runs with a density spectrum.
\label{T1}}
\tablewidth{0pt}
\tablehead{
     \colhead{Run} & \colhead{$\delta$} & \colhead{$l$ [pc]} 
& \colhead{$k_{\rm max}/k_{\rm min}$} \\
}
\startdata
SP20a    &$  3/2  $&$  250  $&$  20  $ \\
SP4a    &$  3/2  $&$  250  $&$  4  $ \\
SP20b    &$  3/2  $&$  25  $&$  20  $ \\
SP4b    &$  3/2  $&$   25  $&$  4  $ \\
SP20c    &$  1/2  $&$   250  $&$  20  $ \\
SP4c    &$  1/2  $&$   250  $&$  4  $ \\
SP20d    &$  1/2  $&$   25  $&$  20  $ \\
SP4d    &$  1/2  $&$   25  $&$  4  $ \\
SP20e    &$  1/2  $&$   2.5  $&$  20  $ \\
SP4e    &$  1/2  $&$   2.5  $&$  4  $ \\
\enddata
\end{deluxetable}

\clearpage

\begin{deluxetable}{cccccccc}
\tablecaption{Summary of runs with forcing.
\label{tab:T2}}
\tablewidth{0pt}
\tablehead{
     \colhead{Run} & \colhead{$l$ [pc]} & \colhead{mesh points} & \colhead{$f_{0}$}
& \colhead{$\left<e\right>$} 
& \colhead{${\mathcal{M}}_{\rm rms}$} & \colhead{$\tau_{\rm cond}$}  \\
}
\startdata
$1$           &$  3  $&$  1200  $&$  0.01 $&$ 0.985 $&$ 0.03 $&$  11.9, 12.1  $\\
$2$           &$  3  $&$  1200  $&$  0.03 $&$ 0.990 $&$ 0.07 $&$   8.4, 10.85 $\\
$3$           &$  3  $&$  1200  $&$  0.09 $&$ 1.095 $&$ 0.18 $&$   6.15, 7.8  $\\
$4$           &$  3  $&$  1200  $&$  0.27 $&$ 1.880 $&$ 0.30 $&$   3.95, 4.6, 6.40  $\\
$5$           &$  3  $&$  1200  $&$  0.36 $&$ 2.450 $&$ 0.30 $&$  >16.0 $\\
$6$           &$  3  $&$  2400  $&$  0.36 $&$ 2.530 $&$ 0.30 $&$  >12.0 $\\
$7$           &$  3  $&$  2400  $&$  0.48 $&$ 2.610 $&$ 0.30 $&$  >14.0 $\\
$8$           &$  3  $&$   600  $&$  0.72 $&$ 3.070 $&$ 0.40 $&$  >26.0 $\\
$9$           &$  3  $&$  3600  $&$  1.20 $&$ 3.300 $&$ 0.65 $&$  \infty $\\
$10$          &$ 10  $&$  1200  $&$  0.01 $&$ 0.980 $&$ 0.027 $&$  10.80, 11.20  $\\
$11$          &$ 10  $&$  2400  $&$  0.04 $&$ 0.980 $&$ 0.10  $&$   7.30, 7.40  $\\
$12$          &$ 10  $&$  7200  $&$  0.04 $&$ 0.980 $&$ 0.10  $&$   7.45, 8.60  $\\
$13$          &$ 10  $&$  1200  $&$  0.10 $&$ 0.995 $&$ 0.13  $&$   5.1, 5.4, 7.0  $\\
$14$          &$ 10  $&$  7200  $&$  0.18 $&$ 1.035 $&$ 0.25  $&$   3.75, 3.90  $\\
$15$          &$ 10  $&$  1200  $&$  0.24 $&$ 1.135 $&$ 0.30  $&$   4.4, 4.6  $\\
$16$          &$ 10  $&$  1200  $&$  0.72 $&$ 1.945 $&$ 0.60  $&$   1.55, 1.70  $\\
$17$          &$ 100  $&$ 9200  $&$ 0.09  $&$ 0.940 $&$ 0.30  $&$   1.78, 1.90 $\\
\enddata
\end{deluxetable}


\begin{thebibliography}{}

\bibitem[Ballesteros-Paredes et al.(1999a)]{bhv99}
Ballesteros-Paredes, J., Hartmann, L., \& V\'azquez-Semadeni, E. 1999,
ApJ, 532, 353

\bibitem[Balbus(1995)]{bal95} 
Balbus, S. A. 1995, in The physics of star formation and 
early stellar evolution, ed. C. J. Lada \& N. D. Kylafis (Netherlands~:~Kluwer
 Academic Publishers), 328

\bibitem[Balbus \& Soker(1989)]{bal89} Balbus, S. A., \& Soker, N. 1989,
ApJ, 341, 611 

\bibitem[Bania \& Lyon(1980)]{ban80} 
Bania, T. M.,\ \& Lyon, J. G. 1980, ApJ, 239, 173

\bibitem[Blitz \& Shu(1980)]{bli80}
Blitz, L., \&  Shu, F. H. 1980, ApJ, 238, 148

\bibitem[Brandenburg(2001)]{bra01} 
Brandenburg, A. 2001, \apj, 550, 824 

\bibitem[Brinkmann, Massaglia \& M\"{u}ller(1990)]{bri90} 
Brinkmann, W., Massaglia, S., \& M\"{u}ller, E. 1990, 
\aap, 237, 536

\bibitem[Burkert \& Lin(2000)]{bur00} 
Burkert, A., \& Lin, D. N. C. 2000, \apj, 537, 270 


\bibitem[Chiang \& Bregman(1988)]{chi88} 
Chiang, W.-H., \&  Bregman, J. N. 1988, \apj, 328, 427 

\bibitem[Dalgarno \& McCray(1972)]{dal72} 
Dalgarno, A., \&  McCray, R. A. 1972, \araa, 10, 375

\bibitem[David, Bregman \& Seab(1988)]{dav88} 
David, L. P., Bregman, J. N., \& Seab, C. G. 1988, \apj, 329, 66

\bibitem[Dickey et al.(1977)]{dic77} 
Dickey, J.M., Salpeter, E.E., \& Terzian, Y. 1977, \apj, 211, L77

\bibitem[Dickey et al.(2001)]{dic01} 
Dickey, J.M., McClure-Griffiths,
N. M., Stanimirovic, S., Gaensler, B. M., \& Green, A. J. 2001, \apj, 561, 264

\bibitem[Elmegreen(1990)]{elm90} 
Elmegreen, B. G. 1990, in The evolution of the Interstellar Medium,
ed. L. Blitz, Astr.Soc.Pac.Conf.Ser. 12, 247

\bibitem[Ferri\'{e}re(2001)]{fer01} 
Ferri\`ere, K. 2001, Rev. Mod. Phys., 73, 1031

\bibitem[Field(1965)]{fie65} 
Field, G. B. \ 1965, \apj, 142, 531

\bibitem[Field, Goldsmith \& Habing(1969)]{fie69} 
Field, G. B., Goldsmith, D. W., \& Habing, H. J. 1969, \apj,
155, L149

\bibitem[Fitzpatrick \& Spitzer(1997)]{fit97} 
Fitzpatrick, E. L., \& Spitzer, L. Jr. 1997, \apj, 475, 623


\bibitem[Gammie \& Ostriker(1996)]{gam96} 
Gammie, C. F., \& Ostriker, E. C. 1996, \apj, 466, 814

\bibitem[Gazol et al.(2001)]{gaz01} 
Gazol, A., V\'{a}zquez-Semadeni, E.,
S\'{a}nchez-Salcedo, F. J., \& Scalo, J. 2001, \apj, 557, L121
(Paper II)

\bibitem[Goldsmith(1970)]{Gol70} 
Goldsmith, D.W. 1970, \apj, 161, 41

\bibitem[Hattori \& Habe(1990)]{hat90} 
Hattori, M.,\ \& Habe, A. 1990, \mnras, 242, 399

\bibitem[Heiles(2001)]{hei01} 
Heiles, C. 2001, \apj, 551, L105

\bibitem[Hennebelle \& P\'erault(1999)]{hen99} 
Hennebelle, P., \& P\'erault, M. 1999, \aap, 351, 309

\bibitem[Hennebelle \& P\'erault(2000)]{hen00} 
Hennebelle, P., \& P\'erault, M. 2000, \aap, 359, 1124

\bibitem[Hollenbach et al.(2001)]{hol01} 
Hollenbach, D. J., Parravano, A., \&  McKee, C. F. 2001, AAS, 199, 117.06

\bibitem[Hunter(1970)]{hun70} 
Hunter, J. H. 1970, \apj, 161, 451

\bibitem[Hunter(1971)]{hun71} 
Hunter, J. H. 1971, \apj, 166, 453

\bibitem[Kalberla et al.(1985)]{kal85} 
Kalberla, P. M., Schwartz, U. J., \& Goss, W. M. 1985, \aap, 144, 27  

\bibitem[Kang et al.(1990)]{kan90} 
Kang, H., Shapiro, P. R., Fall, S. M., \& Rees, M. J. 1990, \apj, 363, 488

\bibitem[Kang, Lake \& Ryu(2000)]{kan00} 
Kang, H., Lake, G., \& Ryu, D. 2000, Journal of Korean
Astronomical Society, 33, 111

\bibitem[Kim et al.(1998)]{kim98} 
Kim, S., Staveley-Smith, L., Dopita, M. A., Freeman, K. C.,
Sault, R. J., Kesteven, M. J., \& McConnell, D. 1998, \apj, 503, 674

\bibitem[Klein, McKee \& Woods(1995)]{kle95} 
Klein, R.,  \& McKee, C. F., \& Woods, D. T. 1995, in The
Physics of the Interstellar Medium and Intergalactic Medium, ASP Conference 
Series, Vol. 80, ed. A. Ferrara, C. F. McKee, C. Heiles, P. R. Shapiro 
(San Francisco~:~ASP), p.~366 

\bibitem[Kornreich \& Scalo(2000)]{kor00} 
Kornreich, P.\ \& Scalo, J.\ 2000, \apj, 531, 366

\bibitem[Korpi et al.(1999)]{kor99} 
Korpi, M. J., Brandenburg, A., Shukurov, A., Tuominen, I.,
\& Nordlund, A.\ 1999, \apj, 514, L99 

\bibitem[Koyama \& Inutsuka(2000)]{koy00} Koyama, H., \& Inutsuka,
S.-I. 2000, \apj, 532, 980

\bibitem[Koyama \& Inutsuka(2002)]{koy02} Koyama, H., \& Inutsuka,
S.-I. 2002, \apj, 564, L97

\bibitem[Kritsuk \& Norman(2002)]{kri02} 
Kritsuk, A., \& Norman, M.L. 2002, \apj, 569, L127

\bibitem[Lioure \& Chi\`eze(1990)]{lio90} 
Lioure, A.,\ \& Chi\`eze, J.-P.\ 1990, \aap, 235, 379 

\bibitem[Mac Low et al.(1999)]{mac99}
Mac Low, M.-M., Klessen, R. S., Burkert, A., \& Smith M. D. 1998, Phys.
Rev. Letters, 80, 2754 

\bibitem[Malagoli, Rosner \& Fryxell(1990)]{mal90} 
Malagoli, A., Rosner, R., \& Fryxell, B. 1990, \mnras, 247, 367

\bibitem[McKee \& Ostriker(1977)]{mck77} 
McKee, C. F.,\ \& Ostriker, J. P.\ 1977, \apj, 218, 148

\bibitem[Meerson(1996)]{mee96} 
Meerson, V. I. 1996, Rev.~Mod.~Phys., 68, 215

\bibitem[Minter \& Spangler(1996)]{min96} 
Minter, A.H., \& Spangler, S.R. 1996, \apj, 458, 194

\bibitem[Murray \& Lin(1989)]{mur89} 
Murray, S. D., \& Lin, D. N. C. 1989, \apj, 339, 933

\bibitem[Murray \& Lin(1990)]{mur90} 
Murray, S. D., \& Lin, D. N. C. 1990, \apj, 363, 50 

\bibitem[Murray \& Lin(1996)]{mur96} 
Murray, S. D., \& Lin, D. N. C. 1996, \apj, 467, 728 

\bibitem[Murray et al.(1993)]{mur93} Murray, S. D., White, S. D. M.,
Blondin, J. M., \& Lin, D. N. C. 1993, \apj, 407, 588

\bibitem[Norman \& Ferrara(1996)]{nor96} 
Norman, C. A., \& Ferrara, A. 1996, \apj, 467, 280

\bibitem[Ostriker et al.(2001)]{ost01} Ostriker, E. C., Stone, J. M., \& 
Gammie, C. F. 2001, \apj, 546, 980


\bibitem[Padoan \& Nordlund(1999)]{pad99} 
Padoan, P., \& Nordlund, A. 1999, \apj, 526, 279


\bibitem[Parravano(1987)]{par87} 
Parravano, A. 1987, \aap, 172, 280 

\bibitem[Passot, V\'azquez-Semadeni \& Pouquet(1995)]{pas95} 
Passot, T., V\'azquez-Semadeni, E., \& Pouquet, A. 1995, \apj, 
455, 536

\bibitem[Raymond, Cox \& Smith(1976)]{ray76} 
Raymond, J. C., Cox, D. P., \& Smith, B. W. 1976, \apj, 204, 290

\bibitem[Reale et al.(1991)]{rea91}
Reale, F., Rosner, R., Malagoli, A., Peres, G., Serio, S. 1991, \mnras, 251, 379

\bibitem[Rosen \& Bregman(1995)]{ros95} 
Rosen, A., \& Bregman, J. N. 1995, \apj, 440, 634

\bibitem[S\'anchez-Salcedo \& Brandenburg(2001)]{san01} 
S\'anchez-Salcedo, F. J., \& Brandenburg, A. 2001, \mnras, 322, 67


\bibitem[Schwarz, McCray \& Stein(1972)]{sch72} 
Schwarz, J., McCray, R., \& Stein, R. F. 1972, \apj, 175, 673

\bibitem[Shu(1992)]{shu92} Shu, F. H. 1992, The Physics of
Astrophysics, vol.~II, (Sausalito, University Science Books)


\bibitem[Spitzer \& Fitzpatrick(1995)]{spi95} 
Spitzer, L. Jr., \& Fitzpatrick, E. L. 1995, \apj, 445, 196

\bibitem[Stanimirovic \& Lazarian(2001)]{sta01} Stanimirovic, S., \&
Lazarian, A. 2001, \apj, 551, L53 

\bibitem[V\'azquez-Semadeni, Passot \& Pouquet(1995)]{vaz95} 
V\'azquez-Semadeni, E., Passot, T., \& Pouquet, A. 
1995, \apj, 441, 702

\bibitem[V\'azquez-Semadeni, Passot \& Pouquet(1996)]{vaz96} 
V\'azquez-Semadeni, E., Passot, T., \& Pouquet, A. 
1996, \apj, 473, 881

\bibitem[V\'azquez-Semadeni \&  Passot(1999)]{vaz99} 
V\'azquez-Semadeni, E. \& Passot, T. 1999, in Interstellar
Turbulence, eds.\ J. Franco and A. Carrami\~nana (Cambridge:
Univ. Press), p.\ 223

\bibitem[V\'azquez-Semadeni, Gazol \& Scalo(2000)]{vaz00} 
V\'azquez-Semadeni, E., Gazol, A., \& Scalo, J. 2000, \apj, 540, 271 (Paper I)

\bibitem[V\'azquez-Semadeni(2002)]{vaz02} V\'azquez-Semadeni, E. 2002, in
Seeing Through the Dust. The Detection of HI and the Exploration of the
ISM in Galaxies, eds.~R.~ Taylor, T.~ Landecker \& T.~ Willis (San
Francisco: ASP), in press

\bibitem[V\'azquez-Semadeni et al.(2002)]{vaz02b} 
V\'azquez-Semadeni, E., Gazol, A., Passot, T., 
\& S\'anchez-Salcedo, F. J. 2002, submitted
to Simulations of Magnetohydrodynamic Turbulence in Astrophysics,
eds.~T.~Passot \& E.~Falgarone (Springer) (Paper III)

\bibitem[Vietri et al.(1997)]{vie97} Vietri, M., Ferrara, A., \&
Miniati, F. 1997, \apj, 483, 262

\bibitem[Wada, Spaans \& Kim(2000)]{wad00} 
Wada, K., Spaans, M., \& Kim, S. 2000, \apj, 540, 797

\bibitem[Wolfire, Hollenbach \& McKee(1995)]{wol95} 
Wolfire, M. G., Hollenbach, D., \& McKee, C. F., Tielens, A. G. G. M. \&
Bakes, E. L. O. 1995, \apj, 443, 152

\bibitem[Zeldovich \& Pikel'ner(1969)]{zel69} 
Zeldovich, Y., \& Pikel'ner, S. 1969, JETP, 29, 170

\end{thebibliography}
\end{document}